\documentclass[prd,aps,floats,onecolumn,twoside,preprintnumbers,
superscriptaddress,floatfix,nofootinbib,showpacs]{revtex4}
\usepackage{graphicx,colordvi,pstricks,rotating,amsmath}
\usepackage{amssymb,epstopdf,bbm}
\usepackage{slashed}
\usepackage[T1]{fontenc}
\usepackage{epsfig}

\begin{document}

\title{UV/IR mixing in noncommutative SU(N) Yang-Mills theory}

\author{C.P. Martin}
\affiliation{Universidad Complutense de Madrid (UCM), Departamento de Física Teórica and IPARCOS, Facultad de Ciencias Físicas, 28040 Madrid, Spain}
\email{carmelop@fis.ucm.es}
\author{Josip Trampeti\'{c}}
\affiliation{Ru\dj er Bo\v{s}kovi\'{c} Institute, Division of Experimental Physics, Bijeni\v{c}ka 54, 10000 Zagreb, Croatia}
\email{josip.trampetic@irb.hr}
\affiliation{Max-Planck-Institut f\"ur Physik, (Werner-Heisenberg-Institut), F\"ohringer Ring 6, 
D-80805 M\"unchen, Germany}
\email{trampeti@mppmu.mpg.de}
\author{Jiangyang You}
\affiliation{Ru\dj er Bo\v{s}kovi\'{c} Institute, Division of Physical Chemistry, Bijeni\v{c}ka 54, 10000 Zagreb, Croatia}
\email{jiangyang.you@irb.hr}

\newcommand{\tr}{\hbox{tr}}
\def\BOX{\mathord{\vbox{\hrule\hbox{\vrule\hskip 3pt\vbox{\vskip
3pt\vskip 3pt}\hskip 3pt\vrule}\hrule}\hskip 1pt}}

\date{today}

\begin{abstract}
We show that there are one-loop  IR singularities arising from  UV/IR mixing  in noncommutative SU(N) Yang-Mills theory defined by means of the $\theta$-exact Seiberg-Witten map. This is in spite of the fact that there are no ordinary U(1) gauge fields in the theory and this is at variance with the noncommutative U(N) case, where the two-point part of the effective action involving the ordinary SU(N) fields do not suffer from those one-loop IR singularities.
\end{abstract}

\maketitle 

\section{Introduction}
Underlying mathematical structures
\cite{Kontsevich:1997vb,Madore:2000en,Jurco:2000fb,Jurco:2001rq,Jurco:2001kp,arXiv0711.2965B,arXiv0909.4259B}
in the noncommutative (NC) quantum field theories (NCQFT)
lead to observable consequences for the low energy physics, realised first by
the perturbative loop computation proposed by Filk~\cite{Filk:1996dm}.
Second is famous example of running of the coupling constant in
the U(1) NCGFT in the star($\star$)-product formalism \cite{Martin:1999aq}.

Since commutative gauge symmetry and its deformed noncommutative gauge symmetry suppose
describe the same physical system --open strings ending on a stack of Dp-branes with
a constant magnetic field on it--, therefore they are expected to be equivalent.
However due to the NC spacetime there are number of principal problems:
Commutative local gauge transformations for the D-brane effective action
do not commute with $\star$-products. Also it is important to note that the introduction
of $\star$-products induces field operator ordering ambiguities
and breaks ordinary gauge invariance in the naive sense, as well as there is also a question of the charge quantisation.

The  problems above were solved by celebrated Seiberg-Witten (SW) maps \cite{Seiberg:1999vs} between ordinary and the noncommutative quantities/fields. SW map promote not only the noncommutative fields
and composite nonlocal operators of the commutative fields, but also
the NC gauge transformations as the composite operators of the commutative gauge fields and gauge transformations
\cite{Okawa:2001mv,Brace:2001fj,Barnich:2002tz,Cerchiai:2002ss,Barnich:2002pb,Banerjee:2004rs,Martin:2012aw,Horvat:2011qn,Trampetic:2015zma}.
This method, referred as SW map or enveloping algebra approach avoids both
the gauge group and the charge quantization problem issues \cite{Horvat:2011qn}.
Thus the NC deformed QFT's can be defined for arbitrary gauge group representations \cite{Martin:2012aw},
and building semi-realistic NC deformed particle physics models are made easier.

One of the particularly intriguing quantum effect of the spacetime noncommuttivity is the exhibition of fascinating dynamics due to the ultraviolet/infrared  phenomenon, named as the UV/IR mixing, an interrelation between short and long distance scales absent in ordinary quantum field theory, in which the ultraviolet and infrared degrees of freedom are mixed in a way similar to what is expected to happen in a theory of Quantum Gravity. In NC theory without SW maps \cite{Bigatti:1999iz,Minwalla:1999px,VanRaamsdonk:2000rr,VanRaamsdonk:2001jd,Hayakawa:1999yt,Hayakawa:1999zf,Hayakawa:2000zi,Matusis:2000jf,Ruiz:2000hu,Khoze:2000sy,Armoni:2003va,Ferrari:2004ex} it was shown for the first time how UV short distance effects, could alter the IR dynamics, thus becoming known as the celebrated UV/IR mixing. With the SW map included \cite{Zeiner:2007,Schupp:2008fs,Horvat:2011bs,Horvat:2011qg,Horvat:2013rga,Horvat:2015aca} we have found the same UV/IR property. Note here that UV/IR mixing shows up also in the NC deformation parameter exact Scalar $\phi^4$ theories and on the $\kappa$-Minkowski and Snyder manifolds, respectively \cite{Grosse:2005iz,Meljanac:2011cs,Meljanac:2017grw,Meljanac:2017jyk}.

UV/IR mixing as an important quantum effect, connects the noncommutative field theories with holography
in a model-independent way \cite{Cohen:1998zx,Horvat:2010km}.

Using the weak gravity conjecture (WGC) \cite{Li:2006jja,Huang:2006tz} and scalar fields results from \cite{Palti:2017elp}, the very notion of UV/IR mixing was implemented into the idea of
scalar fields WGC in the recent L\"ust and Palti article \cite{Lust:2017wrl}; it manifests itself as a form of hierarchical UV/IR mixing and is tied to the interaction
between the weak gravity conjecture and nonlocal (possibly noncommutative) gauge operators. Namely in the scalar field theory, the mass of the scalar is far from ultra-violet scale.
Thus the form of hierarchical UV/IR mixing restricts the mass of the scalar field to a IR scale deep below the UV scale which is associated to quantum gravity.
Additionally one can think about naturalness of the scalar mass in a way that the UV physics knows nothing about the theory in the deep IR. So instead of introducing the higher dimensions and new symmetries one might hope that connecting the deep-low IR and far-high UV could solve the naturalness problem \cite{Craig:2019zbn,Koren:2020biu}.

In the literature, all above notions with UV/IR mixing connections are considered as possible windows to Quantum Gravity.

A chief characteristic of noncommutative U(N) theory is that only the U(1) part of
the two-point function for the gauge field exhibits UV/IR mixing
at the one-loop level. Indeed, the SU(N) part of that two-point function does not
get any contribution from one-loop nonplanar diagrams.
This raises the question of whether noncommutative SU(N) gauge theories experience this UV/IR mixing,
for there is by construction no fundamental U(1) degrees of freedom in them.
The purpose of this paper is to answer this question, which as we shall see demands to carry out lengthy computations.
Let us recall that to formulate Yang-Mills theories on noncommutative
spacetime for arbitrary gauge groups in arbitrary representations one
has to use the noncommutative framework put forward in references \cite{Madore:2000en, Jurco:2001rq, Calmet:2001na, Aschieri:2002mc}. In this formalism the noncommutative gauge fields and
noncommutative gauge transformations are defined solely in terms of
the ordinary counterparts by using the Seiberg-Witten map, which in
this case take values in the universal enveloping algebra of the Lie
algebra of the gauge group.
Thus, any --unitary-- representation of the Lie algebra is admissible; but, then, one needs to address  the problem of too many degrees of freedom, since all coefficient functions of the monomials in the generators could a priori be physical fields. The solution furnished in [2,4,15] is that those coefficient fields are not all independent: they are  functions of the correct number of ordinary gauge fields via SW maps. Thus, for semisimple Lie algebras, no fundamental U(1) degrees of freedom occur in the noncommutative gauge theory.
That the Seiberg-Witten map in question
exists for any compact gauge group –for SU(N), in particular– in any
representation has been shown in reference  \cite{Barnich:2002pb} by using BRST
techniques, thus proving that the Seiberg-Witten exists not only for
the U(N) group in the fundamental, antifundamental or bifundamental
representations. The noncommutative Yang-Mills action for arbitrary
compact gauge groups has been introduced in references \cite{Madore:2000en, Jurco:2001rq, Calmet:2001na, Aschieri:2002mc}  and fully studied in reference  \cite{Barnich:2002pb}. It is by using so-called
enveloping algebra formalism that we have just mentioned that the
noncommutative Yang-Mills theory for SU(N) in the fundamental
representation is formulated –see reference \cite{Calmet:2001na}– without the need of
introducing the interacting U(1) fundamental degrees of freedom which occur in the noncommutative U(N) case.

Using the enveloping algebra formalism we have just made discussed, the background-field and path integral methods, from the classical Yang-Mills (YM) action we shall first construct the BRST exact noncommutative effective action $S_{BRSTexact}$, with the NC fields spanned on the Moyal manifold and being expressed in terms of ordinary YM fields by means of $\theta$-exact SW maps. Let stress that the quantization of the theory will be carried out by integrating in the path integral over the ordinary SU(N) gauge fields and ghosts --this is the SU(N) analogue of quantization method succesfully used in \cite{Horvat:2011qg, Horvat:2013rga, Martin:2016zon} for the U(1) case.  Then, we investigate the UV/IR mixing phenomena in the gauge sector and for a Dirac fermions in the fundamental representation of SU(N). Finally we have to state clearly that this theory is not really the pure Moyal NCYM, but a class of Moyal deformed gauge theory which is different, and not yet completely understood.

This article is on the line of our previous works
\cite{Martin:2016zon,Martin:2016hji,Martin:2016saw,Martin:2017nhg}, and it is organized as follows.
In the next section we introduce classical and the background-field effective actions.
We compute one-loop two-point functions by using DeWitt method
\cite{DeWitt1,Kallosh:1974yh,DeWitt2,DeWitt3} and in Section three
prove existence of the gauge independent UV/IR mixing phenomena. Section four is devoted to
the fundamental Dirac fermions in the framework of the noncommutative QCD.
Conclusion is given in fifth Section, while details of $\theta$-exact SW maps
for all relevant fields and details of divergent integral solutions are given in Appendices A and B, respectively.


\section{The classical action and the background-field effective action}

Let $a_\mu=a^a_{\mu}\,T^a$, in terms of component fields $a^a_\mu$, be an ordinary  gauge field, with $T^a$ being  the generators of SU(N) in the fundamental representation, normalized so that
Tr$(T^a T^b)=\delta^{ab}$. The symbol $A_{\mu}$ will denote the noncommutative gauge field defined in terms of the previous SU(N) ordinary gauge field $a_\mu$ by means of the Seiberg-Witten map \cite{Seiberg:1999vs}.

The classical action, $S_{cl}[A]$, of our noncommutative SU(N) YM theory reads

\begin{equation}
S_{cl}[A(a_{\lambda})]=-\frac{1}{4}\int d^4 x\,{\rm Tr}\: F_{\mu\nu}[A(a_{\lambda})]\star F^{\mu\nu}[A(a_{\lambda})],
\label{classact}
\end{equation}
where $F_{\mu\nu}[A(a_{\lambda})]=\partial_{\mu} A_{\nu}(a_{\lambda})-\partial_{\nu} A_{\mu}(a_{\lambda})-i g [ A_{\mu}(a_{\lambda}),  A_{\nu}(a_{\lambda})]_{\star}${\footnote{ We stress here that $F_{\mu\nu}[A(a_{\lambda})]$ in (\ref{classact}) takes values in the universal enveloping algebra of the SU(N) Lie algebra in the fundamental representation, thus demands the use of the formalism put forward in references \cite{Madore:2000en, Jurco:2001rq, Calmet:2001na, Aschieri:2002mc, Barnich:2002pb}. }}, with star($\star$)-product being the Moyal-Weyl one, and self-evident notations for $\star$-commutator. Important to note is that the dynamical field variable in $S_{cl}[A(a_{\lambda})]$ is $a_\lambda$. For details of SW maps see Appendix A.

To quantize the classical theory whose action is, $S_{cl}[A(a_{\lambda})]$, in (\ref{classact}), we shall use the background field method \cite{Khoze:2000sy} and, thus, we shall split $a_{\mu}$ as follows
\begin{equation}
a_{\mu}=b_{\mu}\,+\,q_{\mu},
\label{splitting}
\end{equation}
where $b_{\mu}=b^a_{\mu}\, T^a$ denotes the background field and $q_{\mu}=q^a_{\mu}\, T^a$ stands for the quantum field --the field to be integrated over in the path integral.

Since the noncommutative gauge field, $A_{\mu}$, is a function of $a_{\mu}$ --$A_{\mu}=A_{\mu}(a_{\lambda})$--, the splitting in (\ref{splitting}) gives rise to the following splitting of $A_{\mu}$
\begin{equation}
A_{\mu}(b+q)=B_{\mu}(b)\,+\,Q_{\mu}(b,q).
\label{NCsplitting}
\end{equation}
Notice that $A_{\mu}(b+q)$ is obtained from $A_{\mu}(a_{\lambda})$ by replacing $a_\mu$ with $b_{\mu}\,+\,q_{\mu}$ in it.
Obviously, $B_{\mu}(b)$ is given by the Seiberg-Witten map applied to the ordinary SU(N) gauge field $b_\mu$. Mark that  $Q_{\mu}(b,q)$ is a function of both $b_\mu$ and $q_\mu$.

The background field effective action $\Gamma[b]$ is given by the following equation
\begin{equation}
e^{i\Gamma[b]}= \int {\cal D}q^a_\mu {\cal D}F^a {\cal D}c^a {\cal D}\bar{C}^a  \, e^{i{\cal S} -i \int d^4 x\,
q^a_{\mu}(x)\frac{\delta \Gamma[b]}{\delta b^a_{\mu}(x)}},
\label{effectiveaction}
\end{equation}
with
\begin{equation}
{\cal S} = S_{cl}[A(b+q)]+ S_{BRSTexact}.
\label{fullaction}
\end{equation}
Classical action, $S_{cl}[A(b+q)]$, is defined in  (\ref{classact}) and $S_{BRSTexact}$ stands for the gauge-fixing terms including the ghost's contributions. To render the sum $S_{cl}[A(b+q)]+  S_{gf+gh}$ as simple as possible we shall choose the following Feynman type of gauge fixing ($S_{gf}$)
and ghost ($S_{gh}$) actions sum $S_{gf+gh}$
\begin{equation}
 S_{BRSTexact} = s\int d^4 x\,{\rm Tr}\;\Big( \bar{C}\star\big(\frac{1}{2}F+D_\mu [B]Q^{\mu}(b,q) \big)\Big)
\end{equation}
where $s$ denotes de ordinary BRST operator, and $\bar{C}=\bar{C}^a\, T^a$ is the antighost, while $F=F^a\,T^a$ Lautrup-Nakanishi auxiliary field. As noted the $S_{BRSTexact}$ is BRST-exact.

The BRST operator $s$ acts on the ordinary and noncommutative fields, respectively,
\begin{equation}
s b_{\mu}= 0,\quad s q_\mu = \partial_\mu c-i g[b_\mu + q_\mu,c],\quad sc =igcc,\quad s \bar{C} = F,\quad s F=0,
\label{ordinaryBRS}
\end{equation}
with $c= c^a\, T^a$ denoting ordinary ghost field. The notion of Seiberg-Witten map implies the action of $s$ on the
noncommutative fields $B_\mu (b)$, $Q_\mu(b,q)$ and $C(b+q,c)$ is the following
\begin{equation}
s B_{\mu}(b)= 0,\quad s Q_\mu(b,q) = s_{NC}Q_\mu (b,q),\quad s C(b+q,c) =s_{NC} C(b+q,c),
\label{ordinaryBRS}
\end{equation}
where $s_{NC}$ is the noncommutative BRST operator whose action on the noncommutative field thus runs as
follows
\begin{equation}
\begin{array}{l}
{s_{NC} B_{\mu}(b)= 0,}\\[4pt]
{ s_{NC}Q_\mu(b,q)= D_\mu [B+Q]C(b+q,c)=\partial_\mu C(b+q,c)-i g[B_\mu +  Q_\mu ,C]_{\star}\,,}\\[4pt]
{s_{NC}C(b+q,c) =igC(b+q,c)\star C(b+q,c).}
\label{NCBRS}
\end{array}
\end{equation}
By  construction, we also have
$ s_{NC} \bar{C}= s \bar{C} = F$ and  $s_{NC} F=s F=0$.

Using  the definitions above, one gets
\begin{equation}
 S_{BRSTexact} = \int d^4 x\,{\rm Tr}\;\Big(\frac{1}{2}F\star F + F\star D_\mu [B]Q^{\mu}(b,q)- \bar{C}\star D_\mu [B]D^\mu [B+Q]C(b+q,c)\Big).
 \label{theF}
\end{equation}

\section{The two-point contribution to the effective action at one-loop}

The final purpose of this section is the computation of one-loop contribution to the effective action
$\Gamma[b]$, which has been defined in (\ref{effectiveaction}).
We shall begin by expanding in powers of $q_\mu$ and $b_\mu$ the action, ${\cal S}$, in (\ref{fullaction}) and, then, dropping the terms with more than
two $b_\mu$'s and more than two $g$'s --$g$ being the coupling constant.

\subsection{Removing $O(g^3)$ terms from the path integral}

Let us first integrate out the field $F^a$ in the path integral in (\ref{effectiveaction}) by taking advantage of (\ref{theF}). Thus, one obtains
\begin{equation}
e^{i\Gamma[b]}= \int {\cal D}q^a_\mu {\cal D}c^a {\cal D}\bar{C}^a  \, e^{i S -i \int d^4 x\,
q^a_{\mu}(x)\frac{\delta \Gamma[b]}{\delta b^a_{\mu}(x)}},
\label{effectiveactionnew}
\end{equation}
where $S$ is the sum
\begin{equation}
S = S_{cl}[A(b+q)]+ S_{gf}+S_{gh};
\label{fullactiongf}
\end{equation}
while actions $S_{gf}$ and $S_{gh}$ being given by
\begin{equation}
\begin{array}{l}
{S_{gf} =
-\frac{1}{2} \int d^4 x\,{\rm Tr}\; (D_\mu [B]Q^{\mu}(b,q)\star D_\nu [B]Q^{\nu}(b,q))}\\[4pt]
{\phantom{S_{gf} = }+\frac{1}{2 N}\int d^4 x\,({\rm Tr} D_\mu[B]Q^{\mu}(b,q))\star ({\rm Tr}  D_\nu[B]Q^{\nu}(b,q)),}\\[8pt]
{S_{gh}= -\int d^4 x\,{\rm Tr}\;   (\bar{C}\star D_\mu [B]D^\mu [B+Q]C(b+q,c)).}
\label{Sgaugefix}
\end{array}
\end{equation}
To obtain above $S_{gf}$ we have to use the following SU(N) generators identity
\begin{equation}
\sum_a(T^a)^{i_1}_{\bar{j}_1}(T^a)^{i_2}_{\bar{j}_2}= \delta^{i_1}_{\bar{j}_2}\delta^{i_2}_{\bar{j}_1}-\frac{1}{N}\delta^{i_1}_{\bar{j}_1}\delta^{i_2}_{\bar{j}_2}.
\label{closurerel}
\end{equation}

We start with definitions
\begin{equation}
\tilde{q}^a_\mu= {\rm Tr}\, (T^a Q_\mu(b,q)), \;{\rm and}\; \tilde{c}^a= {\rm Tr}\, (T^a C(b,q)),
\label{3.5}
\end{equation}
where the Seiberg-Witten maps $Q_\mu(b,q)$ and $C(b,q)$ are fully discussed in Appendix A --see (\ref{thehatAS})
and (\ref{theCbc}). 
Next, in the path integral (\ref{effectiveactionnew}), we shall make the following variable change
\begin{equation}
q^a_\mu\rightarrow \tilde{q}^a_\mu,\quad c^a\rightarrow \tilde{c}^a,
\label{changeofvariables}
\end{equation}
for thus we shall remove from the path integral the lengthy interaction terms due to contributions to the classical action coming from the SU(N) part
--see (\ref{theUNcombination})-- of the Seiberg-Witten maps $Q_\mu(b,q)$ and $C(b,q)$ when expressed as functions of $q_\mu$. This is a much
welcome simplification since, as we shall see below,  we will still have to deal with the interaction terms the U(1) part of the former Seiberg-Witten maps introduce. Notice that the Jacobian of this transformation is trivial in Dimensional Regularization, since all the momentum integrals it involves vanish --for details see \cite{Martin:2016saw}.

Now, let $\Gamma_2 [b]$ be the two-point contribution to $\Gamma[b]$ in (\ref{effectiveactionnew}). The change of variables (\ref{changeofvariables})
and the use of (\ref{theQbq}) and (\ref{thectilde}) from Appendix A, after some laborious algebra, leads to
\begin{equation}
i \Gamma_2 [b] = Ln \int {\cal D}\tilde {q}^a_\mu {\cal D}\tilde{c}^a {\cal D}\bar{C}^a  \; e^{i (S_0+S_1+S_2 + S_{1eom}+S_{2eom}+ S_{0gh}+S_{1gh}+S_{2gh})}+O(g^3),
\label{2pteffective}
\end{equation}
where
\begin{equation}
\begin{array}{l}
{S_0=-\frac{1}{2} \int d^4 x\, {\rm Tr}\,\, \partial_\mu \tilde{q}_\nu \partial^\mu \tilde{q}^{\nu},}\\[8pt]
{S_1=+ig \int d^4 x\, {\rm Tr}\,\, b_\mu[\tilde{q}_{\nu},\partial^\mu \tilde{q}^{\nu}]_{\star}
+ 2ig \int d^4 x\, {\rm Tr}\,\, \partial_\mu b_\nu[\tilde{q}^{\mu}, \tilde{q}^{\nu}]_{\star}\,,}\\[8pt]
{S_2=+\frac{1}{2}g^2\int d^4 x\, {\rm Tr}\,\,[b_{\mu}, \tilde{q}_{\nu}]_{\star}[b^{\mu}, \tilde{q}^{\nu}]_{\star}-
\frac{N}{2}g^2\int d^4 x\,\partial_\mu \hat{A}^{(2)\,0}_\nu (b,\tilde{q})\partial^\mu \hat{A}^{(2)\,0\,\nu} (b,\tilde{q})}\\[4pt]
{\phantom{S_2=}+i g^2 \int d^4 x\, {\rm Tr}\,\,\hat{A}^{(2)}_\mu (b,b)[\tilde{q}_{\nu},\partial^\mu \tilde{q}^{\nu}]_{\star}
+i g^2 \int d^4 x\, {\rm Tr}\,\,\hat{A}^{(2)\,0}_\nu (b,\tilde{q})[\partial_\mu \tilde{q}^{\nu},b^{\mu}]_{\star}}\\[4pt]
{\phantom{S_2=}+i g^2 \int d^4 x\, {\rm Tr}\,\,\partial_\mu\hat{A}^{(2)\,0}_\nu (b,\tilde{q})[b^\mu, \tilde{q}^{\nu}]_{\star}
+g^2 \int d^4 x\, {\rm Tr}\,\, \big\{\partial^2-\partial_\nu\partial^\rho \hat{A}^{(2)}_\rho(b,b)\big\}\hat{A}^{(2)\,0\,\nu}(\tilde{q},\tilde{q})}\\[4pt]
{\phantom{S_2=}+g^2\int d^4 x\, {\rm Tr}\,\,[b_{\mu}, b_\nu]_{\star}[\tilde{q}^{\mu}, \tilde{q}^{\nu}]_{\star}+
2 i g^2 \int d^4 x\, {\rm Tr}\,\,\partial_\mu \hat{A}^{(2)}_\nu (b,b)[\tilde{q}^\mu, \tilde{q}^{\nu}]_{\star}}\\[4pt]
{\phantom{S_2=}+2 i g^2 \int d^4 x\, {\rm Tr}\,\partial_\mu\hat{A}^{(2)\,0}_\nu (b,\hat{q})[b_\nu, \tilde{q}^{\mu}]_{\star}
+\frac{N}{2}g^2\int d^4 x\, \partial_\mu \hat{A}^{(2)\,0\,\mu} (b,\tilde{q})\partial_\nu \hat{A}^{(2)\,0\,\nu} (b,\tilde{q})}\\[4pt]
{\phantom{S_2=}-\frac{1}{2 N}g^2\int d^4 x\, {\rm Tr}\,\,[b_{\mu}, \tilde{q}^{\mu}]_{\star}[b^{\nu}, \tilde{q}^{\nu}]_{\star}-
i g^2\int d^4 x\,{\rm Tr}\,\,\partial_\mu \hat{A}^{(2)\,0\,\mu} (b,\tilde{q})[b_\nu,\tilde{q}^\nu]_{\star}\,,}
\label{S012}
\end{array}
\end{equation}
\begin{equation}
\begin{array}{l}
{S_{1eom}=g \int d^4 x\, {\rm Tr}\,\{\big(\partial^2 b^\nu-\partial^\nu\partial_\rho b^\rho\big)\hat{A}^{(2)}_\nu (\tilde{q},\tilde{q}) \},}\\[8pt]
{S_{2eom}=+g^2\int d^4 x\, {\rm Tr}\,\{\big(\partial^2 b^\nu-\partial^\nu\partial_\rho b^\rho\big)\hat{A}^{(3)}_\nu(\tilde{q},\tilde{q},b)\}}\\[4pt]
{\phantom{S_{2eom}=}-2 g^2 \int d^4 x\, {\rm Tr}\,\{\big(\partial^2 b^\nu-\partial^\nu\partial_\rho b^\rho\big)\hat{A}^{(2)}_\nu (\hat{A}^{(2)\,a}(\tilde{q},b)T^a,\tilde{q}) \}}\\[4pt]
{\phantom{S_{2eom}=}-g^2\int d^4 x\, {\rm Tr}\,\big(E_\nu(b,b)\hat{A}^{(2)}_\nu (\tilde{q},\tilde{q})\big)+
\frac{g^2}{N} \int d^4 x\, {\rm Tr}\,\big(E_\nu(b,b)\big){\rm Tr}\,\big(\hat{A}^{(2)}_\nu (\tilde{q},\tilde{q})\big)
}\\[4pt]
{\quad\quad E_\nu(b,b)=\partial^2 \hat{A}^{(2)}_\nu(b,b)-\partial_\nu\partial^\rho \hat{A}^{(2)}_\rho(b,b)-i\partial^\mu [b_\mu,b_\nu]_{\star}
-i[b^\mu,\partial_\mu b_\nu-\partial_\nu b_\mu]_{\star}\,,}
\label{Seom012}
\end{array}
\end{equation}
and
\begin{equation}
\begin{array}{l}
{S_{0gh}=-\int d^4 x\, {\rm Tr}\,\, \bar{C}\partial^2 \tilde{c},}\\[8pt]
{S_{1gh}=ig \int d^4 x\, {\rm Tr}\,\, b^\mu\{\tilde{c},\partial_\mu \bar{C}\}-ig\int d^4 x\, {\rm Tr}\,\, b^\mu\{\partial_\mu\tilde{c},\bar{C}\},}\\[8pt]
{S_{2gh}=-g^2 \int d^4 x\, {\rm Tr}\,\, [b^\mu,\tilde{c}]_{\star}[b_\mu, \bar{C}]_{\star}+ig^2\int d^4 x\, {\rm Tr}\,\,\bar{C} [b^\mu,\partial_\mu \hat{C}^{(2)\,0}(\tilde{c},b)]_{\star}}\\[4pt]
{\phantom{S_{2gh}=}+ig^2 \int d^4 x\, {\rm Tr}\,\, \hat{A}^{(2)}_\mu(b,b)\{\tilde{c},\partial^\mu\bar{C}\}_{\star}
-ig^2 \int d^4 x\, {\rm Tr}\,\,\hat{A}^{(2)}_\mu(b,b)\{\partial^\mu\tilde{c},\bar{C}\}_{\star}}\\[4pt]
{\phantom{S_{2gh}=}-ig^2\int d^4 x\, {\rm Tr}\,\, \hat{C}^{(2)\,0}(\tilde{c},b)[\partial_\mu\bar{C},b^\mu]_{\star}\,,}
\label{S012gh}
\end{array}
\end{equation}
with self-evident notations for equations of motions ($eom$), and $\star$-anticommutator.
Notice that $S_2$ and $S_{2gh}$ only involve the U(1) part of the Seiberg-Witten maps $Q_\mu(b,q)$ and $C(b,c)$ as defined in (\ref{thehatAS})
and (\ref{theCbc}): the SU(N) parts of those maps have been disposed of  by introducing the SU(N) fields $\tilde{q}$ and $\tilde{c}$--see (\ref{theQbq}) and (\ref{thectilde}) in Appendix A.

Now, let $p^\mu$ and $q^\mu$ be two arbitrary vectors. We shall use the following notation
\begin{equation}
\begin{array}{l}
p\wedge q = \theta^{\mu\nu}p_\mu q_\nu\equiv p_\mu\theta^{\mu\nu} q_\nu\equiv p\theta q
=p\tilde{q},\quad \tilde{q}^\mu= \theta^{\mu\nu} q_\nu,\\[8pt]
{(f\star_t g)(x)=e^{i(p_1+p_2)x} e^{-i\frac{t}{2}p_1 \wedge p_2} f(p_1) g(p_2)}.
\label{tstar}
\end{array}
\end{equation}
Then, $\hat{A}^{(2)\,0}_\nu (b,\tilde{q})$, $\hat{A}^{(2)}_\nu (b,b)$, $\hat{A}^{(2)}_\mu (\tilde{q},\tilde{q})$,
$\hat{A}^{(3)}_\mu (\tilde{q},\tilde{q},b)$, $\hat{A}^{(2)}_\nu(\hat{A}^{(2)\,a}(\tilde{q},b)T^a,\tilde{q})$ 
and $C^{(2)\,0}(\tilde{c},b)$, are given by the following expressions:

\begin{equation}
\begin{array}{l}
{\hat{A}^{(2)\,0}_\mu (b,\tilde{q})=\frac{1}{2 N}\,e^{i(p_1+p_2)x}\big(\frac{e^{-\frac{i}{2}p_1\wedge p_2}-e^{\frac{i}{2}p_1\wedge p_2}}{p_ 1\wedge p_ 2}\big)}\\[3pt]
{\phantom{\hat{A}^{(2)\,0}_\mu (b,\tilde{q})=\frac{1}{2 N}XXXXXX,
}
\times[2 \tilde{p}_{2}^{\mu_1}\delta^{\mu_2}_\mu+ 2\tilde{p}_{1}^{\mu2}\delta^{\mu_1}_\mu -(p_2-p_1)_\mu \theta^{\mu_1 \mu_2}] b^a_{\mu_1}(p_1)\tilde{q}^a_{\mu_2}(p_2),}\\[8pt]
{\hat{A}^{(2)}_\nu (b,b)=\frac{1}{2}\,e^{i(p_1+p_2)x}\big(\frac{e^{-\frac{i}{2}p_1\wedge p_2}-1}{p_ 1\wedge p_ 2}\big)}\\[3pt]
{\phantom{\hat{A}^{(2)}_\nu (b,b)=\frac{1}{2}XXXXXx
}\quad\quad
 \times[2 \tilde{p}_{2}^{\mu_1}\delta^{\mu_2}_\mu+ 2\tilde{p}_{1}^{\mu2}\delta^{\mu_1}_\mu -(p_2-p_1)_\mu \theta^{\mu_1 \mu_2}] b_{\mu_1}(p_1)b_{\mu_2}(p_2),}\\[8pt]
{\hat{A}^{(2)}_\mu (\tilde{q},\tilde{q};t)=
-\frac{i}{4}\,e^{i(p_1+p_2)x}\,\int_0^t ds\, e^{-i\frac{s}{2}p_1\wedge p_ 2}}\\[3pt]
{\phantom{\hat{A}^{(2)}_\mu (\tilde{q},\tilde{q};t)=\frac{i}{4}XXXXx
}\quad\quad
\times[2 \tilde{p}_{2}^{\mu_1}\delta^{\mu_2}_\mu+ 2\tilde{p}_{1}^{\mu2}\delta^{\mu_1}_\mu -(p_2-p_1)_\mu \theta^{\mu_1 \mu_2}] \tilde{q}_{\mu_1}(p_1)\tilde{q}_{\mu_2}(p_2),}\\[8pt]
{\hat{A}^{(2)}_\mu (\tilde{q},\tilde{q})=\hat{A}^{(2)}_\mu (\tilde{q},\tilde{q};t=1),}\\[8pt]
{\hat{A}^{(2)}_\mu (\tilde{q},b;t)=-\frac{i}{4}\,e^{i(p_1+p_2)x} [2 \tilde{p}_{2}^{\mu_1}\delta^{\mu_2}_\mu+ 2\tilde{p}_{1}^{\mu2}\delta^{\mu_1}_\mu -(p_2-p_1)_\mu \theta^{\mu_1 \mu_2}]}\\[4pt]
{\phantom{\hat{A}^{(2)}_\mu (\tilde{q},b;t)=\frac{i}{4}
}\quad
\times\int_0^t ds\,[e^{-i\frac{s}{2}p_1\wedge p_2}b_{\mu_1}(p_1)\tilde{q}_{\mu_2}(p_2)+e^{i\frac{s}{2}p_1\wedge p_2}\tilde{q}_{\mu_2}(p_2)b_{\mu_1}(p_1)],}\\[8pt]
{\hat{A}^{(3)}_\mu (\tilde{q},\tilde{q},b)=
-\frac{1}{4}\theta^{ij}\int_0^1 dt}\\[4pt]
\phantom{XXX}\times\Big[+\{\tilde{q}_i,2\partial_j\hat{A}^{(2)}_\mu (\tilde{q},b;t)-\partial_\mu\hat{A}^{(2)}_j (\tilde{q},b;t)\}_{\star_t}
{+\{\hat{A}^{(2)}_i(\tilde{q},b;t),2\partial_j\tilde{q}_\mu-\partial_\mu\tilde{q}_j\}_{\star_t}}\\[4pt]
{\phantom{XXXXx,}
+\{b_i,2\partial_j\hat{A}^{(2)}_\mu (\tilde{q},\tilde{q};t)-\partial_\mu\hat{A}^{(2)}_j (\tilde{q},\tilde{q};t)\}_{\star_t}
+\{\hat{A}^{(2)}_i(\tilde{q},\tilde{q};t),2\partial_j b_\mu-\partial_\mu b_j\}_{\star_t}}\\[4pt]
{\phantom{XX\hat{A}^{(3)}_\mu (\tilde{q},\tilde{q},b)=-\frac{1}{4}\theta^{ij}\int_0^1 dt\Big[}
-i\{\tilde{q}_i,[\tilde{q}_j,b_\mu]_{\star_t}\}_{\star_t}
-i\{\tilde{q}_i,[b_j,\tilde{q}_\mu]_{\star_t}\}_{\star_t}
-i\{b_i,[\tilde{q}_j,\tilde{q}_\mu]_{\star_t}\}_{\star_t}\Big]},\\[8pt]
{\hat{A}^{(2)}_\nu(\hat{A}^{(2)\,a}(\tilde{q},b)T^a,\tilde{q})=
-\frac{1}{16}\int\frac{d^4\!p_2}{(2\pi)^4}\frac{d^4\!p_3}{(2\pi)^4}\frac{d^4\!p_4}{(2\pi)^4}\quad e^{i(p_2+p_3+p_4)x}}\\[3pt]
{\times[2\tilde{p}_2^{\mu_1}\delta^{\mu_2}_\nu+2(\tilde{p}_3+\tilde{p}_4)^{\mu_2}\delta^{\mu_1}_\nu+(p_3+p_4-p_2)_\nu\theta^{\mu_1\mu_2}]
[2\tilde{p}_4^{\mu_3}\delta^{\mu_4}_{\mu_1}+2\tilde{p}_3^{\mu_4}\delta^{\mu_3}_{\mu_1}+(p_3-p_4)_{\mu_1}\theta^{\mu_3\mu_4}]}\\[4pt]
{\qquad\quad\quad\quad\times\Big\{\int_0^1 dt\int_0^1 ds\,e^{-i\frac{t}{2}(p_3+p_4)\wedge p_2}e^{-i\frac{s}{2}p_3\wedge p_4}
[{\rm Tr}\,(b_{\mu_3}(p_3)\tilde{q}_{\mu_4}(p_4)T^b)]T^b\tilde{q}_{\mu_2}(p_2)}\\[3pt]
{\qquad\quad\quad\quad\phantom{\Big\{}+\int_0^1 dt\int_0^1 ds\,e^{-i\frac{t}{2}(p_3+p_4)\wedge p_2}e^{+i\frac{s}{2}p_3\wedge p_4}
[{\rm Tr}\,(\tilde{q}_{\mu_4}(p_4)b_{\mu_3}(p_3)T^b)]T^b\tilde{q}_{\mu_2}(p_2)}\\[3pt]
{\qquad\quad\quad\quad\phantom{\Big\{}+\int_0^1 dt\int_0^1 ds\,e^{+i\frac{t}{2}(p_3+p_4)\wedge p_2}e^{-i\frac{s}{2}p_3\wedge p_4}
\tilde{q}_{\mu_2}(p_2)[{\rm Tr}\,(b_{\mu_3}(p_3)\tilde{q}_{\mu_4}(p_4)T^b)]T^b}\\[3pt]
{\qquad\quad\quad\quad\phantom{\Big\{}+\int_0^1 dt\int_0^1 ds\,e^{+i\frac{t}{2}(p_3+p_4)\wedge p_2}e^{+i\frac{s}{2}p_3\wedge p_4}
\tilde{q}_{\mu_2}(p_2)[{\rm Tr}\,(\tilde{q}_{\mu_4}(p_4)b_{\mu_3}(p_3)T^b)]T^b\Big\},}\\[8pt]
{C^{(2)\,0}(\tilde{c},b)=-\frac{1}{2N}\theta^{ij}\,e^{i(p_1+p_2)x}\,p_{1i}\,\big(\frac{e^{-\frac{i}{2}p_1\wedge p_2}-e^{\frac{i}{2}p_1\wedge p_2}}{p_ 1\wedge p_ 2}\big)\tilde{c}^a(p_1)b_j(p_2).}
\label{SWmaps}
\end{array}
\end{equation}

\subsection{The one-loop contribution to the two-point function}

By evaluating the order $g^2$ contribution to the right hand side of eq. (\ref{2pteffective}),
we shall obtain the one-loop contribution to $\Gamma_2[b]$,
the two-point bit of the effective action. The result that we obtained, with help of integral basis given in Appendix B and \cite{Grozin:2000cm,Martin:2016zon}, runs thus
\begin{equation}
\begin{array}{l}
{\Gamma_2[b]=\frac{g^2}{N} \int\frac{d^4 p}{(2\pi)^4}\,{\rm Tr}\,\big[b_{\mu}(p)b_\nu(-p)\big]}\\[3pt]
{\phantom{.}\times\Big\{\frac{1}{16 \pi^2}\frac{1}{(\tilde{p}^2)^3}\Big[\frac{16}{3}p^2\theta^{\mu i}\tilde{p}_i \theta^{\nu j}\tilde{p}_j-
\frac{8}{3}\tilde{p}^2(\theta^{\mu i}\tilde{p}_i p^\nu+ \theta^{\nu j}\tilde{p}_j p^\mu)-\frac{32}{3}(\tilde{p}^2)^2\eta^{\mu\nu}+
\frac{32}{3}\tilde{p}^2\tilde{p}^\mu \tilde{p}^\nu \Big]}\\[3pt]
{\phantom{x}-\frac{1}{\pi^2}\frac{\tilde{p}^\mu \tilde{p}^\nu}{(\tilde{p}^2)^2}\int_0^1 dx\,\Big[x(1-x)p^2\tilde{p}^2K_2(\sqrt{x(1-x)p^2\tilde{p}^2})\Big]\Big\}}\\[3pt]
\phantom{x}+{\frac{g^2}{16 \pi^2} \,\int\frac{d^4 p}{(2\pi)^4} {\rm Tr}\,\big[b_{\mu}(p)\big(p^2\eta^{\mu\nu}-p^\mu p^\nu\big)b_\nu(-p)\big]
\Big\{-\frac{11}{3}\big(N-\frac{2}{N}\big)\Big(\frac{1}{\epsilon}+ Ln\frac{p^2}{4\pi\mu^2}+\gamma-1\Big)}\\[3pt]
{\phantom{\frac{g^2}{16 \pi^2}(N-\frac{2}{N}) \,{\rm Tr}XXXXXXXXXX}
+\frac{4}{N}\int_0^1 dx\, (3+2 x)\,K_0(\sqrt{x(1-x)p^2\tilde{p}^2})\Big\}}\\[3pt]
{+\;\Gamma_2[b]^{(eom0)}+\;({\rm 2-loop\;order})},
\label{quiteacomputation}
\end{array}
\end{equation}
where $\Gamma_2[b]^{(eom0)}$, which obviously vanishes upon imposing the equations of motion, contains all the contributions on which integral
\begin{equation}
\int d^4 x\,q^a_{\mu}(x)\frac{\delta \Gamma[b]}{\delta b^a_{\mu}(x)},
\label{3.13}
\end{equation}
in (\ref{effectiveaction}), is involved. That is, $\Gamma_2[b]^{(eom0)}$ is the sum of all the contributions which involve
either $S_{1eom}$ or $S_{2oem}$ from (\ref{2pteffective}),
and it is given by the following long expression
\begin{equation}
\begin{array}{l}
{\Gamma_2[b]^{(eom0)}=}\\[3pt]
{+\frac{g^2}{16 \pi^2}\int\frac{d^4 p}{(2\pi)^4}\,
 {\rm Tr}\,[b_{\mu}(p)(p^2\eta^{\mu\nu}-p^\mu p^\nu)b_\nu(-p)]}\\[3pt]
{\phantom{\Gamma_2[b]^{(eom0)}=}
\times\Big\{ N\, [\frac{15}{\epsilon}+ 15 Ln(\frac{p^2}{4\pi\mu^2})+\frac{1}{15} Ln(p^2\tilde{p}^2)+60 Ln\,2 -15\gamma -30] }\\[3pt]
{\phantom{\Gamma_2[b]^{(eom0)}=}\phantom{\Big\{N\, [\frac{15}{\epsilon}+ 15 Ln\frac{p^2}{4\pi\mu^2}}+\frac{1}{N}\,\big[16\big(-\frac{1}{\epsilon}
-Ln\frac{p^2}{4\pi\mu^2}+\gamma+2\big)
-34 Ln\, 2\big]\Big\}}\\[3pt]
{-\frac{1}{N}\,\frac{g^2}{16 \pi^2}\int\frac{d^4 p}{(2\pi)^4}\, {\rm Tr}\,[b_{\mu}(p)(p^2\delta^{\mu}_{\phantom{\mu}\rho}-p^\mu p_\rho)b_\nu(-p)]}\\[4pt]
{\phantom{\Gamma_2[b]^{(eom0)}}\times\Big\{\big[4\big(-\frac{1}{\epsilon}-Ln\frac{p^2}{4\pi\mu^2}-\gamma+2\big)+8\int_0^1 dx\,K_0(\sqrt{x(1-x)p^2\tilde{p}^2})\big]\frac{\tilde{p}_\rho\tilde{p}_\nu}{\tilde{p}^2}}\\[4pt]
{\phantom{\Gamma_2[b]^{(eom0)}=-\frac{1}{N}\,\frac{g^2}{16 \pi^2} Tr[b_{\mu}(p)(p^2\quad\quad}+
\frac{1}{(\tilde{p}^2)^3}(
\frac{28}{3}\tilde{p}^2 \theta^{\rho\sigma}\theta_{\sigma}^{\phantom{\sigma}\nu}+
\frac{84}{3}\tilde{p}^\rho \theta^{\nu i}\theta_{i}^{\phantom{i}\lambda}\tilde{p}_\lambda)\Big\}}\\[3pt]
{+\frac{g^2 N}{16\pi^2}\int\frac{d^4 p}{(2\pi)^4}\, {\rm Tr}\,[b_{\mu}(p)(p^2\delta^{\mu}_{\phantom{\mu}\rho}-p^\mu p_\rho)(p^2\delta_{\sigma}^{\phantom{\sigma}n}-p_\sigma p^\nu)b_\nu(-p)]}\\[3pt]
{\quad\quad\times\Big\{\frac{\eta^{\rho\sigma}}{\tilde{p}^2}[\frac{3}{2}(\theta^{ij}\theta_{ji}){\cal C}_1- \frac{\tilde{p}^2}{p^2}{\cal C}_2-
\frac{\theta^{\lambda i}\tilde{p}_i\theta_{\lambda}^{\phantom{\lambda} j}\tilde{p}_j}{\tilde{p}^2}{\cal C}_4
-\frac{1}{2}{\cal B}_1(\tilde{p}^2)^2]+\frac{\tilde{p}_\rho\tilde{p}_\sigma}{\tilde{p}^2}[{\cal B}_1\tilde{p}^2+\frac{1}{p^2}{\cal C}_2+\frac{1}{2}\theta^{ij}\theta_{ji}{\cal C}_4]}\\[3pt]
{\phantom{\{\eta^{\rho\sigma}[\frac{3}{2}(\theta^{ij}\theta_{ji}){\cal C}_1+ \frac{\tilde{p}^2}{p^2}{\cal C}_2-
\frac{\theta^{\lambda i}\tilde{p}_i\theta_{\lambda}^{\phantom{\lambda} }\tilde{p}_j}{}}
-3\frac{\theta^{\rho\,i}\,\theta_{i}^{\phantom{i}\sigma}}{\tilde{p}^2}{\cal C}_1+
\frac{\theta^{\rho\, i}\tilde{p}_i\,\theta^{\sigma\,j}\tilde{p}_j}{(\tilde{p}^2)^2}{\cal C}_4
-\frac{\theta^{\rho\lambda}\theta_{\lambda}^{\phantom{\lambda}i}\tilde{p}_i\;\tilde{p}^\sigma+
\theta^{\sigma\lambda}\theta_{\lambda}^{\phantom{\lambda}i}\tilde{p}_i\;\tilde{p}^\rho}{(\tilde{p}^2)^2}{\cal C}_4\Big\}}\\[3pt]
{-\frac{g^2}{N}\int\frac{d^4 p}{(2\pi)^4}\, {\rm Tr}\,[b_{\mu}(p)(p^2\delta^{\mu}_{\phantom{\mu}\rho}-p^\mu p_\rho)(p^2\delta_{\sigma}^{\phantom{\sigma}\nu}-p_\sigma p^\nu)b_\nu(-p)]}\\[3pt]
{\quad\quad\times\Big\{\frac{\eta^{\rho\sigma}}{\tilde{p}^2}[3(\theta^{ij}\theta_{ji})({\cal C}_1+\frac{1}{2}\tilde{{\cal C}}_1)- \frac{\tilde{p}^2}{p^2}(2{\cal C}_ 2+\tilde{{\cal C}}_2)
-\frac{\theta^{\lambda i}\tilde{p}_i\theta_{\lambda}^{\phantom{\lambda} j}\tilde{p}_j}{\tilde{p}^2}(2{\cal C}_ 4+\tilde{{\cal C}}_4)}\\[3pt]
{\quad\qquad-({\cal B}_ 1 +\tilde{{\cal B}}_1+\frac{1}{2}\tilde{{\cal A}})(\tilde{p}^2)^2]+\frac{\tilde{p}_\rho\tilde{p}_\sigma}{\tilde{p}^2}[2{\cal B}_1\tilde{p}^2
+\tilde{{\cal B}}_1\tilde{p}^2+\frac{1}{p^2}({\cal C}_2+\tilde{{\cal C}}_2)+\frac{1}{2}\theta^{ij}\theta_{ji}({\cal C}_4+\frac{1}{2}\tilde{{\cal C}}_4])}\\[3pt]
{\qquad\qquad-3\frac{\theta^{\rho\,i}\,\theta_{i}^{\phantom{i}\sigma}}{\tilde{p}^2}(2{\cal C}_1+\tilde{{\cal C}}_1)+
\frac{\theta^{\rho\,i}\tilde{p}_i\,\theta_{\sigma}^{\phantom{\sigma}j}\tilde{p}_j}{(\tilde{p}^2)^2}
(2{\cal C}_4+\tilde{{\cal C}}_4)
-\frac{\theta^{\rho\lambda}\theta_{\lambda}^{\phantom{\lambda}i}\tilde{p}_i\;\tilde{p}^\sigma+ \theta^{\sigma\lambda}\theta_{\lambda}^{\phantom{\lambda}i}\tilde{p}_i\;\tilde{p}^\rho }{(\tilde{p}^2)^2}
({\cal C}_4+\frac{1}{2}\tilde{{\cal C}}_4)\Big\}}\\[3pt]
{-\frac{i}{2}\frac{g^2 N}{16\pi^2}\int\frac{d^4 p}{(2\pi)^4}\, {\rm Tr}\,[b_{\sigma}(p)(p^2\delta^{\sigma}_{\phantom{\sigma}\rho}-p^\sigma p_\rho)f_{\mu\nu}(-p)](4\frac{\theta^{\mu\,i}\tilde{p}_i\eta^{\rho\nu}}{\tilde{p}^2}+2\theta^{\mu\nu}\frac{\tilde{p}^\rho}{\tilde{p}^2})}\\[3pt]
{\qquad\qquad\quad\times\big[2\big(-\frac{1}{\epsilon}-Ln\frac{p^2}{4\pi\mu^2}-\gamma+2\big)-4\int_0^1 dx\,K_0(\frac{\sqrt{x(1-x)p^2\tilde{p}^2}}{2})\big]}\\[3pt]
{+\frac{i}{N}\frac{g^2}{16\pi^2}\int\frac{d^4 p}{(2\pi)^4}\, {\rm Tr}\,[b_{\sigma}(p)(p^2\delta^{\sigma}_{\phantom{\sigma}\rho}-p^\sigma p_\rho)f_{\mu\nu}(-p)](4\frac{\theta^{\mu\,i}\tilde{p}_i\eta^{\rho\nu}}{\tilde{p}^2}+2\theta^{\mu\nu}\frac{\tilde{p}^\rho}{\tilde{p}^2})}\\[3pt]
{\qquad\times\big[2\big(-\frac{1}{\epsilon}-Ln\frac{p^2}{4\pi\mu^2}-\gamma+2\big)-4\int_0^1 dx\,\big(K_0(\sqrt{x(1-x)p^2\tilde{p}^2})-2 K_0(\frac{\sqrt{x(1-x)p^2\tilde{p}^2}}{2})\big)\big].}
\label{thezeroeom}
\end{array}
\end{equation}
Note that $f_{\mu\nu}(-p)=i(p_\mu b_\nu(-p)-p_\nu b_\mu(-p))$.
${\cal B}_1$, ${\cal C}_1$, ${\cal C}_2$, ${\cal C}_4$, $\tilde{{\cal A}}$, $\tilde{{\cal B}}_1$, $\tilde{{\cal C}}_1$, $\tilde{{\cal C}}_2$ and $\tilde{{\cal C}}_4$ in (\ref{thezeroeom}) read thus:
\begin{equation}
\begin{array}{l}
{{\cal B}_1=-\int_0^1 dx\, x^{-1/2}\int_0^1 dy\,\big[xy(1-y)p^2\tilde{p}^2\big]^{-1}\Big[\frac{\sqrt{xy(1-y)p^2\tilde{p}^2}}{2}
K_1\big(\frac{\sqrt{x(1-x)p^2\tilde{p}^2}}{2}\big)-1\Big]},\\[3pt]
{{\cal C}_1=+\frac{1}{\epsilon}+ Ln\frac{p^2}{4\pi\mu^2}+\gamma+2 Ln(p^2\tilde{p}^2)-8Ln 2+2\gamma+ 3-\int_0^1 dx\, K_0\big(\frac{\sqrt{x(1-x)p^2\tilde{p}^2}}{2}\big)}\\[3pt]
{\phantom{{\cal C_1}\frac{1}{\tilde{p}}}+p^2\tilde{p}^2\int_0^1 dx\, x^{-1/2}\int_0^1 dy\,
\big[xy(1-y)p^2\tilde{p}^2\big]^{-1}\Big[\frac{\sqrt{xy(1-y)p^2\tilde{p}^2}}{2}
K_1\big(\frac{\sqrt{x(1-x)p^2\tilde{p}^2}}{2}\big)-1\Big]},\\[3pt]
{{\cal C}_2=-\frac{9}{\epsilon}-9 Ln\frac{p^2}{4\pi\mu^2}-7\gamma +1- Ln(p^2\tilde{p}^2)+8Ln 2+2\int_0^1 dx\, K_0\big(\frac{\sqrt{x(1-x)p^2\tilde{p}^2}}{2}\big)\Big)}\\[3pt]
{\phantom{{\cal C}_2=}-\frac{3}{4} p^2\tilde{p}^2\int_0^1 dx\, x^{-1/2}\int_0^1 dy\,\big[xy(1-y)p^2\tilde{p}^2\big]^{-1}\Big[\frac{\sqrt{xy(1-y)p^2\tilde{p}^2}}{2}
K_1\big(\frac{\sqrt{x(1-x)p^2\tilde{p}^2}}{2}\big)-1\Big]},\\[3pt]
{{\cal C}_4=-\frac{1}{\epsilon}- Ln\frac{p^2}{4\pi\mu^2}-\gamma -7 -2 Ln(p^2\tilde{p}^2)+8Ln 2+6\int_0^1 dx\, K_0\big(\frac{\sqrt{x(1-x)p^2\tilde{p}^2}}{2}\big)\Big)}\\[3pt]
{\phantom{{\cal C}_4=}-\frac{1}{4}p^2\tilde{p}^2\int_0^1 dx\, x^{-1/2}\int_0^1 dy\,\big[xy(1-y)p^2\tilde{p}^2\big]^{-1}\Big[\frac{\sqrt{xy(1-y)p^2\tilde{p}^2}}{2}K_1\big(\frac{\sqrt{x(1-x)p^2\tilde{p}^2}}{2}\big)-1\Big]},\\[3pt]
{\tilde{{\cal A}}=2\int_0^1 dx\, x^{-1/2}\int_0^1 dy\,\big[xy(1-y)p^2\tilde{p}^2\big]^{-1}}\\[3pt]
{\qquad\qquad\qquad\times\Big\{\sqrt{xy(1-y)p^2\tilde{p}^2}
\Big[K_1\big(\sqrt{x(1-x)p^2\tilde{p}^2}\big)-K_1\big(\frac{\sqrt{x(1-x)p^2\tilde{p}^2}}{2}\big)\Big]+1\Big\}},\\[3pt]
{\tilde{{\cal B}_1}=-\int_0^1 dx\, x^{-1/2}\int_0^1 dy\,\big[xy(1-y)p^2\tilde{p}^2\big]^{-1}}\\[3pt]
{\qquad\qquad\qquad\quad\times\Big\{\sqrt{xy(1-y)p^2\tilde{p}^2}
\Big[K_1\big(\sqrt{x(1-x)p^2\tilde{p}^2}\big)-\frac{1}{2}K_1\big(\frac{\sqrt{x(1-x)p^2\tilde{p}^2}}{2}\big)\Big]\Big\}},\\[3pt]
{\tilde{{\cal C}_1}=-\frac{1}{\epsilon}-Ln\frac{p^2}{4\pi\mu^2}-2 Ln(p^2\tilde{p}^2)+4 Ln 2-3\gamma-3+
\int_0^1 dx\,K_0(\frac{\sqrt{x(1-x)p^2\tilde{p}^2}}{2})}\\[3pt]
{\qquad\;+\frac{p^2\tilde{p}^2}{4}\int_0^1 dx\, x^{-1/2}\int_0^1 dy\,\big[xy(1-y)p^2\tilde{p}^2\big]^{-1}}\\[3pt]
{\qquad\qquad\qquad\times\Big\{\sqrt{xy(1-y)p^2\tilde{p}^2}
\Big[K_1\big(\sqrt{x(1-x)p^2\tilde{p}^2}\big)-\frac{1}{2}K_1\big(\frac{\sqrt{x(1-x)p^2\tilde{p}^2}}{2}\big)\Big]\Big\}},\\[3pt]
{\tilde{{\cal C}}_2=+\frac{9}{\epsilon}+9 Ln\frac{p^2}{4\pi\mu^2}+7\gamma -1+2 Ln(p^2\tilde{p}^2)-4Ln 2}\\[3pt]
{\phantom{\tilde{{\cal C}}_2=}+2\int_0^1 dx\, \Big[K_0\big(\sqrt{x(1-x)p^2\tilde{p}^2}\big)-K_0\big(\frac{\sqrt{x(1-x)p^2\tilde{p}^2}}{2}\big)\Big]}\\[3pt]
{\phantom{{\cal C}_2=}-\frac{3}{4} p^2\tilde{p}^2\int_0^1 dx\, x^{-1/2}\int_0^1 dy\,\big[xy(1-y)p^2\tilde{p}^2\big]^{-1}}\\[3pt]
{\qquad\qquad\qquad\times\Big\{\sqrt{xy(1-y)p^2\tilde{p}^2}
\Big[K_1\big(\sqrt{x(1-x)p^2\tilde{p}^2}\big)-K_1\big(\frac{\sqrt{x(1-x)p^2\tilde{p}^2}}{2}\big)\Big]+1\Big\}},\\[3pt]
{\tilde{{\cal C}}_4=-\frac{1}{\epsilon}- Ln\frac{p^2}{4\pi\mu^2}+\gamma +7+2 Ln(p^2\tilde{p}^2)-12Ln 2}\\[3pt]
{\phantom{\tilde{{\cal C}}_4=}+6\int_0^1 dx\, \Big[K_0\big(\sqrt{x(1-x)p^2\tilde{p}^2}\big)-K_0\big(\frac{\sqrt{x(1-x)p^2\tilde{p}^2}}{2}\big)\Big]}\\[3pt]
{\phantom{{\cal C}_4=}-\frac{1}{4} p^2\tilde{p}^2\int_0^1 dx\, x^{-1/2}\int_0^1 dy\,\big[xy(1-y)p^2\tilde{p}^2\big]^{-1}}\\[3pt]
{\qquad\qquad\qquad\times\Big\{\sqrt{xy(1-y)p^2\tilde{p}^2}
\Big[K_1\big(\sqrt{x(1-x)p^2\tilde{p}^2}\big)-K_1\big(\frac{\sqrt{x(1-x)p^2\tilde{p}^2}}{2}\big)\Big]+1\Big\}.}
\end{array}
\label{3.15}
\end{equation}

\subsection{Gauge independent UV/IR mixing}

It is plane that $\Gamma_{2}[b]$ in (\ref{quiteacomputation}) develops, as a result of UV/IR mixing, IR divergences in the region where $\tilde{p}^\mu=0$ and that these divergences do not survive the large $N$ limit. And yet,
$\Gamma_{2}[b]$ is a gauge-fixing dependent quantity so one may ask whether all these IR singularities are gauge-fixing artifacts. We shall answer this question by putting the background field $b_\mu$ on shell so that our $\Gamma_{2}[b]$ will boil down to the 2-point on-shell DeWitt effective action \cite{DeWitt1,Kallosh:1974yh,DeWitt2,DeWitt3}, for we are computing radiative corrections at order $g^2$. It is known that on-shell DeWitt effective action is a gauge-fixing independent object.

Since we are working at order $g^2$, we only demand the $b_\mu$ be a solution, $b^{(0)}_\mu$, to the free equation of motion to put $b_\mu$ on-shell:
\begin{equation}
\partial^2 b^{(0)}_\mu(x)- \partial_\mu \partial^\nu b^{(0)}_\nu(x)=0.
\label{freeeom}
\end{equation}
Now, any solution to (\ref{freeeom}) is of the form
\begin{equation}
 b^{(0)}_\mu(x) = b^{\bot}_\mu (x)+ \partial_\mu \alpha(x),
\label{bnot}
\end{equation}
where $\alpha(x)$ is an arbitrary function taking values in the Lie algebra of SU(N) and
$b^{\bot}_\mu (x)$ is such that
\begin{equation}
\partial^2 b^{\bot}_\mu (x) = 0,\quad \partial^\mu b^{\bot}_\mu(x)=0.
\label{transverse} 
\end{equation}

Let us replace $b_\mu$ in (\ref{quiteacomputation}) with $b^{(0)}_\mu(x)$ in (\ref{bnot}) to obtain the 2-point on-shell effective action \cite{DeWitt1,Kallosh:1974yh,DeWitt2,DeWitt3}:
\begin{eqnarray}
\Gamma_2[b^{(0)}]&=&\frac{g^2}{N}\int\frac{d^4 p}{(2\pi)^4}\, {\rm Tr}\,\big[b^{\bot}_{\mu}(p) b^{\bot}_\nu(-p)\big]
\Big\{-\frac{2}{3 \pi^2}\frac{\eta^{\mu\nu}}{\tilde{p}^2}+
\frac{2}{3 \pi^2}\frac{\tilde{p}^\mu \tilde{p}^\nu}{(\tilde{p}^2)^2}
\label{DeWitt}\\
&-&\frac{1}{\pi^2}\frac{\tilde{p}^\mu \tilde{p}^\nu}{(\tilde{p}^2)^2}\int_0^1 dx\,[x(1-x)p^2\tilde{p}^2K_2(\sqrt{x(1-x)p^2\tilde{p}^2})]\Big\}+ O(g^3).
\nonumber
\end{eqnarray}
Notice that $b^{\bot}_{\mu}(p)$ above is the Fourier transform of the $b^{\bot}_{\mu}(x)$ satisfying the conditions in (\ref{transverse}) and that is
related to $b^{\bot}_{\mu}(p)$ by a gauge transformation. Notice that the right-hand side of equation (\ref{DeWitt}) is invariant under arbitrary on-shell gauge transformations. The latter as defined thus
\begin{equation}
b^{\bot}_{\mu}(p)\rightarrow b^{\bot}_{\mu}(p)+ p_\mu \beta(p),\quad p^2 \beta(p) =0.
\label{3.21}
\end{equation}

Finally, $\Gamma_2[b^{(0)}]$ above develops quadratic IR singularity, in the limit $\tilde{p}^\mu\to0$, that is given by
\begin{equation}
\frac{g^2}{N}\frac{-2}{ 3\pi^2\tilde{p}^2}\Big\{\eta^{\mu\nu}+
2\frac{\tilde{p}^\mu \tilde{p}^\nu}{\tilde{p}^2}\Big\}{\rm Tr}\,\big[b^{\bot}_{\mu}(p) b^{\bot}_\nu(-p)\big],
\label{IRsingularity}
\end{equation}
which goes away in the large $N$ limit.

Note that obtained (\ref{IRsingularity}) has completely different Lie algebra structure with respect to the famous IR singularity in the one-loop two-point function of the ordinary gauge field of noncommutative U(N) case
\cite{VanRaamsdonk:2001jd},
\begin{equation}
{\frac{2g^2}{\pi^2}}\frac{\tilde{p}^\mu \tilde{p}^\nu}{(\tilde{p}^2)^2}\big[{\rm Tr}\,b^{\bot}_{\mu}(p)\big]\big[{\rm Tr}\, b^{\bot}_\nu(-p)\big].
\label{UNIRsingularity}
\end{equation}

\subsection{The UV/IR mixing phenomenon}

Let us show now that the IR singularity in (\ref{IRsingularity}) which occur when $\tilde{p}^\mu=0$ is a consequence
of the UV/IR mixing phenomenon and not the result of the Seiberg-Witten maps in (\ref{SWmaps}) having denominators which vanish for
specific values of the momenta. We shall display below each and everyone of the contributions to the path integral in (\ref{2pteffective}) which yield (\ref{DeWitt}).

Let us begin with the following definition
\begin{equation}
\big\langle {\cal O}(\tilde{q}^a_\mu,\tilde{c}^a,\bar{C}^a)\big\rangle_{0}=\int {\cal D}\tilde {q}^a_\mu {\cal D}\tilde{c}^a {\cal D}\bar{C}^a  \,
e^{i (S_0+ S_{0gh})}\;{\cal O}(\tilde{q}^a_\mu,\tilde{c}^a,\bar{C}^a),
\end{equation}
where $S_0$ and $S_{0gh}$ are given in (\ref{S012}) and (\ref{S012gh}), respectively. Then,
\begin{equation}
-ig^2\frac{N}{2}\Big\langle\,\partial^{\mu}\hat{A}^{(2)\,0}_\mu(b^{(0)},\tilde{q})(x) \partial^{\nu}\hat{A}^{(2)\,0}_\nu (b^{(0)},\tilde{q})(x)\Big\rangle_{0},
\label{3.23}
\end{equation}
gives rise to the following contribution
\begin{eqnarray}
&-&\frac{g^2}{2N} {\rm Tr}\,\big(b^{(0)}_{\,\mu}(p) b^{(0)}_{\,\nu}(-p)\big)\tilde{p}^2\int\frac{d^4 k}{(2\pi)^4} \frac{k^\mu k^\nu}{k^2 (k\tilde{p})^2}(e^{ik\tilde p}+e^{-ik\tilde p}-2)
\label{IR1}\\
&=& i\frac{g^2}{N}\frac{1}{16\pi^2} {\rm Tr}\big(b^{(0)}_{\,\mu}(p) b^{(0)}_{\,\nu}(-p)\big)
\Big\{\frac{4}{3}\Big(\frac{\eta^{\mu\nu}}{\tilde{p}^2}-4
\frac{\tilde{p}^\mu\tilde{p}^\nu}{(\tilde{p}^2)^2}\Big)\Big\}\,+\,O(\epsilon),
\nonumber
\end{eqnarray}
with $D=4+2\epsilon$. Now the factor
\begin{equation}
\frac{(e^{ik\tilde p}+e^{-ik\tilde p}-2)}{(k\tilde{p})^2},
\label{3.25}
\end{equation}
tends to a constant as $k\tilde{p}$ goes to zero. Hence, the singularity at $\tilde{p}^2=0$ of the right hand side of (\ref{IR1}) is a consequence of the fact that, when $\tilde{p}^\mu\neq 0$  in the integral over $k$ exponentials $e^{\pm ik\tilde p}$ kills the UV divergent behaviour of the rest of the integrand, rendering a finite result. Of course, these UV divergent behaviour resurfaces in the form of a IR divergence as $\tilde{p}^2\rightarrow 0$.
This is precisely celebrated UV/IR mixing of the U(N) noncommutative gauge field theories, discovered first in \cite{Minwalla:1999px,Hayakawa:1999yt,Hayakawa:1999zf,Hayakawa:2000zi,VanRaamsdonk:2000rr,Matusis:2000jf}.

The contribution to (\ref{DeWitt}) coming from
\begin{equation}
-ig^2\frac{N}{2}\Big\langle\,\partial^{\mu}\hat{A}^{(2)\,0\,\nu}(b^{(0)},\tilde{q})(x) \partial_{\mu}\hat{A}^{(2)\,0}_\nu (b^{(0)},\tilde{q})(x)\Big\rangle_{0},
\label{3.26}
\end{equation}
reads
\begin{eqnarray}
&&\frac{3g^2}{2N}
 {\rm Tr}\big(b^{(0)}_{\,\mu}(p) b^{(0)}_{\,\nu}(-p)\big)(\theta^{\mu j_1}\theta^{\nu j_2}+\theta^{\mu j_2}\theta^{\nu j_1})p^\lambda p_{j_2}
\int\frac{d^4 k}{(2\pi)^4} \frac{k_\lambda k_{j_1}}{k^2 (k\tilde{p})^2}(e^{ik\tilde p}+e^{-ik\tilde p}-2)
\nonumber\\
&&= i\frac{g^2}{N}\frac{1}{16\pi^2} {\rm Tr}\,\big(b^{(0)}_{\,\mu}(p) b^{(0)}_{\,\nu}(-p)\big)\Big\{-8
\frac{\tilde{p}^\mu\tilde{p}^\nu}{(\tilde{p}^2)^2}\Big\}\,+\,O(\epsilon).
\label{IR2}
\end{eqnarray}
In the above equation (\ref{IR2}) we meet the very same type of loop integral --the integral over $k$--,
so the IR divergent behaviour as $\tilde{p}^\mu$ goes to zero that occurs on the right hand side of (\ref{IR2}) has the UV/IR mixing origin
that we discuss in the paragraph below (\ref{IR1}).

Let us now deal with
\begin{equation}
-g^2\Big\langle\,{\rm Tr}\,\hat{A}^{(2)\,0}_{\nu}(b^{(0)},\tilde{q})(x) [\partial_{\mu}\tilde{q}^\nu,b^{(0)\,\mu}]_{\star}(x)\Big\rangle_{0}.
\label{3.28}
\end{equation}
It can be shown that the previous expression is equal to
\begin{eqnarray}
&&\frac{g^2}{2N}3 {\rm Tr}\,\big(b^{(0)}_{\,\mu}(p) b^{(0)}_{\,\nu}(-p)\big)\tilde{p}^\mu \int\frac{d^4 k}{(2\pi)^4}\frac{k^\nu}{k^2 (k\tilde{p})}(e^{ik\tilde p}+e^{-ik\tilde p}-2)
\label{IR3}\\
&&= i\frac{g^2}{N}\frac{1}{16\pi^2} {\rm Tr}\,\big(b^{(0)}_{\,\mu}(p) b^{(0)}_{\,\nu}(-p)\big)\Big\{12
\frac{\tilde{p}^\mu\tilde{p}^\nu}{(\tilde{p}^2)^2}\Big\}\,+\,O(\epsilon).
\nonumber
\end{eqnarray}
Notice that
\begin{equation}
\frac{(e^{ik\tilde p}+e^{-ik\tilde p}-2)}{k\tilde{p}},
\label{3.30}
\end{equation}
approaches zero as $\tilde{p}^\mu$ goes to zero. Hence, the vanishing, when $\tilde{p}^\mu\rightarrow 0$, of the denominator of the integral over $k$
in (\ref{IR3}) has no bearing on the IR divergence at $\tilde{p}^\mu=0$ that occurs on the right hand side of (\ref{IR3}). Indeed, again, this IR divergence
at $\tilde{p}^\mu=0$ occurs as a consequence of the fact that $e^{\pm ik\tilde p}$ cuts-off the UV divergent behaviour of the integral over $k$ we have just mentioned:  clearly here  $\tilde p$ acts as a cut-off giving rise to the UV/IR mixing phenomenon.

The loop integral --the integral over $k$-- which occurs in (\ref{IR3}) is also the only responsible for the IR singularity at $\tilde{p}^\mu=0$
of the following contributions to $\Gamma_2[b^{(0)}]$ in (\ref{DeWitt}):
\begin{equation}
\begin{array}{l}
{-g^2\big\langle\,{\rm Tr}\,\partial_\mu\hat{A}^{(2)\,0}_{\nu}(b^{(0)},\tilde{q})(x) [b^{(0)\,\mu},\tilde{q}^\nu]_{\star}(x)\big\rangle_{0}}\\[4pt]
{\phantom{g\langle\,{\rm Tr}\,\partial_\mu\hat{A}^{(2)\,0}_{\nu}(b^{(0)} }
=i\frac{g^2}{N}\frac{1}{16\pi^2} {\rm Tr}\,\big(b^{(0)}_{\,\mu}(p) b^{(0)}_{\,\nu}(-p)\big)\Big\{12
\frac{\tilde{p}^\mu\tilde{p}^\nu}{(\tilde{p}^2)^2}\Big\}\,+\,\text{additional terms},}\\[8pt]
{-2g^2\big\langle\,{\rm Tr}\,\hat{A}^{(2)\,0}_{\nu}(b^{(0)},\tilde{q})(x) [\partial_\mu b^{(0)\,\nu},\tilde{q}^\mu]_{\star}(x)\big\rangle_{0}}\\[4pt]
{\phantom{g\langle\,{\rm Tr}\,\partial_\mu\hat{A}^{(2)\,0}_{\nu}(b^{(0)} }
=i\frac{g^2}{N}\frac{1}{16\pi^2} {\rm Tr}\,\big(b^{(0)}_{\,\mu}(p) b^{(0)}_{\,\nu}(-p)\big)\Big\{8
\frac{\tilde{p}^\mu\tilde{p}^\nu}{(\tilde{p}^2)^2}\Big\}\,+\,\text{additional terms},}\\[8pt]
{+g^2\big\langle\,{\rm Tr}\,\partial_\mu\hat{A}^{(2)\,0}_{\mu}(b^{(0)},\tilde{q})(x) [b^{(0)}_{\,\nu},\tilde{q}^\nu]_{\star}(x)\big\rangle_{0}}\\[4pt]
{\phantom{g\langle\,{\rm Tr}\,\partial_\mu\hat{A}^{(2)\,0}_{\nu}(b^{(0)} }
=i\frac{g^2}{N}\frac{1}{16\pi^2} {\rm Tr}\,\big(b^{(0)}_{\,\mu}(p) b^{(0)}_{\,\nu}(-p)\big)\Big\{-8
\frac{\tilde{p}^\mu\tilde{p}^\nu}{(\tilde{p}^2)^2}\Big\}\,+\,\text{additional terms}.}
\label{IR4}
\end{array}
\end{equation}

There remains to discuss the origin of the singularity at $\tilde{p}^\mu=0$ of the following contributions to $\Gamma_2[b^{(0)}]$ in (\ref{DeWitt}):
\begin{equation}
\begin{array}{l}
{-i\frac{g^2}{2N}\big\langle\,{\rm Tr}\, [b^{(0)}_\mu,\tilde{q}_\nu]_{\star}(x)[b^{(0)\,\mu},\tilde{q}^\nu]_{\star}(x)\big\rangle_{0}=
\frac{ig^2}{16\pi^2N} {\rm Tr}\,\big(b^{(0)}_{\,\mu}(p) b^{(0)}_{\,\nu}(-p)\big)\Big\{-4
\frac{\eta^{\mu\nu}}{\tilde{p}^2}\Big\},}\\[10pt]
{-i\frac{g^2}{2}\big\langle\,{\rm Tr}\, [b^{(0)}_\mu,\tilde{q}^\mu]_{\star}(x)[b^{(0)}_\nu,\tilde{q}^\nu]_{\star}(x)\big\rangle_{0}=
\frac{ig^2}{16\pi^2N} {\rm Tr}\,\big(b^{(0)\,\mu}(p) b^{(0)}_{\,\nu}(-p)\big)\Big\{-16
\frac{\eta^{\mu\nu}}{\tilde{p}^2}\Big\},}\\[10pt]
{-i\frac{g^2}{2}\big\langle\Big(\,{\rm Tr}\, b^{(0)}_\mu[\tilde{q}_\nu,\partial^\mu\tilde{q}_\nu]_{\star}\Big)^2\big\rangle_{0}
-2ig^2\big\langle\Big(\,{\rm Tr}\, \partial_\mu b^{(0)}_\nu[\tilde{q}^\mu,\tilde{q}_\nu]_{\star}\Big)^2\big\rangle_{0}}\\[4pt]
{\qquad-i\frac{g^2}{2}\big\langle\Big(\,{\rm Tr}\, b^{(0)}_\mu\{\partial^\mu\tilde{c},\bar{C}\}_{\star}-
-\,{\rm Tr}\, b^{(0)}_\mu\{\tilde{c},\partial^\mu\bar{C}\}_{\star}\Big)^2\big\rangle_{0}}\\[4pt]
{=\frac{ig^2}{16\pi^2N} {\rm Tr}\,\big(b^{(0)}_{\,\mu}(p) b^{(0)}_{\,\nu}(-p)\big)16\big\{\frac{\eta^{\mu\nu}}{\tilde{p}^2}
-\frac{\tilde{p}^\mu \tilde{p}^\nu}{(\tilde{p}^2)^2}\int_0^1 dx\,[x(1-x)p^2\tilde{p}^2K_2(\sqrt{x(1-x)p^2\tilde{p}^2})]\Big\}, }\\[10pt]
{-ig^2\big\langle\,{\rm Tr}\, [b^{(0)}_\mu,\tilde{c}]_{\star}(x)[b^{(0)\,\mu},\bar{C}]_{\star}\big\rangle_{0}=
-\frac{ig^2}{16\pi^2N} {\rm Tr}\,\big(b^{(0)}_{\,\mu}(p) b^{(0)}_{\,\nu}(-p)\big)\Big\{8\frac{\eta^{\mu\nu}}{\tilde{p}^2}\Big\}.}\\[8pt]
\label{IR5}
\end{array}
\end{equation}
The singular behaviour at $\tilde{p}^\mu=0$ in the expression above comes uniquely from various types of integrals over $k$, all being given in Appendix B:
\begin{equation}
\,\frac{1}{k^2}\,e^{ik\tilde{p}},\;\;\,\frac{(2 k+p)_\mu(2 k+p)_\nu}{k^2(k+p)^2}\,e^{ik\tilde{p}},\;\;......
 \label{3.33}
\end{equation}
The UV/IR mixing phenomenon that the latter integrals bring about has been amply discussed in the literature already \cite{Minwalla:1999px, Hayakawa:1999yt}. Notice that one obtains (\ref{DeWitt}) by adding the right hand sides of (\ref{IR1}), (\ref{IR2}), (\ref{IR3}), (\ref{IR4}) and (\ref{IR5}).

Some final comments are in order. The U(1)  part of the Seiberg-Witten map $Q_\mu(b,q)$ as defined in (\ref{theUNcombination}), contributes to the singularity at $\tilde{p}_\mu=0$ due to UV/IR mixing
--see (\ref{IR1}--\ref{IR4}). The UV/IR mixing effect the SU(N) bit of $Q_\mu(b,q)$ which
is involved in, is obtained by using the field $\tilde{q}_\mu$, --see (\ref{IR5}).

\section{Adding Dirac fermions in the fundamental: Noncommutative QCD}

We shall show in this section that the inclusion of a Dirac fermion transforming under the fundamental representation of SU(N) does not change
the one-loop IR singular behaviour of the two point function of $b_\mu(x)$, $\Gamma_2[b]$, that we have unveiled in the previous section.

Let $\psi(x)=\psi_j(x)$, $j=1,...,N$--the SU(N) index, be an ordinary Dirac fermion transforming under the fundamental representation of SU(N).
Let $\Psi(x)=\Psi(a,\psi)_j(x)$, with $a_\mu(x)=b_\mu(x)+q_\mu(x)$, be the noncommutative fermion field obtained from ordinary fields $\psi_j(x)$
by using the Seiberg-Witten map. Then the classical action  $S_{Dirac}$, of $\Psi(a,\psi)_j(x)$ coupled to
the noncommutative gauge field $A_\mu(x)=B_\mu(b)(x)+Q_\mu(b,q)(x)$, reads
\begin{equation}
S_{Dirac}\,=\,\, \bar{\Psi}(a,\psi)\big(i\slashed D\,[A]-m\big)\Psi(a,\psi),
\label{4.1}
\end{equation}
where $\bar{\Psi}(a,\psi)$ is the Dirac conjugate of $\Psi(x)$, and $\slashed D\,[A]=i(\slashed\partial-g\slashed A\star)$.

The one-loop contribution, $\Gamma_{Dirac}[b]$, to the background field effective action $\Gamma[b]$ due to the noncommutative Dirac field $\Psi(a,\psi)$ is given by the path integral with respect to the ordinary fields $\psi,\;\bar\psi$
\begin{equation}
e^{i\Gamma_{Dirac}[b]}=\int {\cal D}\psi{\cal D}{\bar{\psi}}\;e^{i\, \bar{\Psi}(b,\psi)\big(i\slashed D\,[B]-m\big)\Psi(b,\psi)}.
\label{Diracpath}
\end{equation}

Now, since $\Psi(a,\psi)_j$ is a four component spinor, one can make the following change of variables
\begin{equation}
\psi\rightarrow \Psi=\Psi(b,\psi).
\label{4.3}
\end{equation}
in the path integral (\ref{Diracpath}) and obtain:
\begin{equation}
e^{i\Gamma_{Dirac}[b]}=\int {\cal D}\Psi{\cal D}{\bar{\Psi}}\;J_F(b)\bar{J}_F(b)\, e^{i\, \bar{\Psi}(b,\psi)\big(i\slashed D\,[B]-m\big)\Psi(b,\psi)}
\label{Diracpathtwo}
\end{equation}
where $J_F(b)$ and $\bar{J}_F(b)$ are the appropriate Jacobians:
\begin{equation}
J_F[b]=Det\,\frac{\delta\Psi(b,\psi)_i(x)}{\delta\psi_j(y)}\quad\text{and}\quad\bar{J}_F[b]=Det\,
\frac{\delta\bar{\Psi}(b,\psi)_i(x)}{\delta\bar{\psi}_j(y)}.
\label{4.5}
\end{equation}

As with the Seiberg-Witten map for the gauge field, a SW map defining the noncommutative field $\Psi(b,\psi)$ can be obtained by solving the following ``evolution'' problem:
\begin{eqnarray}
\frac{d\phantom{t}}{dt}\Psi(b,\psi;t)&=&-\frac{g}{2}\theta^{\mu\nu}B_\mu(b;t)\star_t\partial_\nu\Psi(b,\psi;t) +\frac{i}{4}g^2 B_\mu(b;t)\star_t B_\mu(b;t)\star_t\Psi(b,\psi;t),
\nonumber\\
\Psi_\mu(b;t=0)&=&\psi,
\label{SWpsi}
\end{eqnarray}
where $B_\mu(b,t)$ is given by the ``evolution'' equation discussed in Appendix A.

The ``evolution'' problem in (\ref{SWpsi}) can be solved by expanding in powers of $g$ --see Ref.\cite{Martin:2012aw} for details
-- so that one may show that
\begin{eqnarray}
\frac{\delta\Psi(b,\psi)(x)}{\delta \psi(y)}&=&\mathbb{I}\delta(x-y)+ \sum_{n=2}^{\infty}\,g^{n-1}\,\int\prod_{i=1}^n\,e^{i\sum_{i=1}^{n-1}p_i x} e^{ip_n(x-y)}
\label{thematrixJ}\\
&\times&{\cal F}^{\mu_1\mu_2\cdot \mu_{n-1}}(p_1,p_2,...,p_{n-1};p_n;\theta)\, b_{\mu_1}(p_1)b_{\mu_2}(p_2)\cdots b_{\mu_{n-1}}(p_{n-1}).
\nonumber
\end{eqnarray}

It can be seen that ${\cal F}^{\mu_1\mu_2\cdot \mu_{n-1}}(p_1,p_2,...,p_{n-1};p_n;\theta)$ in (\ref{thematrixJ}) is a
linear combination of functions of the type
\begin{equation}
{\cal P}(p_1,....,p_n){\cal Q}(p_i\wedge p_j),
\label{4.8}
\end{equation}
where ${\cal P}(p_1,....,p_n)$ is a polynomial of the momenta $p_i$'s and ${\cal Q}(p_i\wedge p_j)$ only depends on
$p_i\wedge p_j=p_{\mu\,i}\theta^{\mu\nu}p_{\nu\,j}$, $i,j=1,...,n$. This fact leads to the conclusion that all the
loop integrals involved in the evaluation of
\begin{equation}
Det\,\frac{\delta\Psi(b,\psi)_i(x)}{\delta\psi_j(y)}
\label{4.9}
\end{equation}
vanish in dimensional regularization, and, hence,
\begin{equation}
J_F[b]=Det\,\frac{\delta\Psi(b,\psi)_i(x)}{\delta\psi_j(y)}=1.
\label{trivialdet}
\end{equation}
The detailed derivation of (\ref{trivialdet}) is  analogous to the one carried out in Appendices B and C of Ref. \cite{Martin:2016saw}.

Clearly, in dimensional regularization we also have
\begin{equation}
\bar{J}_F[b]=Det\,\frac{\delta\bar{\Psi}(b,\psi)_i(x)}{\delta\bar{\psi}_j(y)}=1.
\label{trivialdetbar}
\end{equation}

Let us introduce the following notation:
\begin{equation}
\langle\cdots\rangle_{f0}=\int {\cal D}\Psi{\cal D}{\bar{\Psi}}\,\cdots\, e^{i\, \bar{\Psi}(i\slashed\partial-m)\Psi.}
\label{4.12}
\end{equation}

Substituting (\ref{trivialdet}) and (\ref{trivialdetbar}) in (\ref{Diracpathtwo}), one readily obtains, in dimensional regularization, the following one-loop result
\begin{eqnarray}
i\Gamma_{Dirac}(b)&=& Ln \sum_{n=0}^{\infty}\frac{(-i g)^n}{n!}\,\langle\big(\; \bar{\Psi}\gamma^\mu B_\mu(b)\star\Psi\big)^n\rangle_{f0}
\label{oneloopfe}\\
&=&-\sum\limits_{n=2}^\infty\,\frac{g^n}{n}\,\prod\limits_{i=1}^n\; {\rm tr} \big(B_{\mu_1} (b)(x_1)\cdots B_{\mu_n}(b)(x_n) \big)\,\Gamma^{\mu_1\cdots\mu_n}(x_1,...x_n),
\nonumber
\end{eqnarray}
where
\begin{eqnarray}
&&{\Gamma^{\mu_1\cdots\mu_n}(x_1,...x_n)=\prod\limits_{i=1}^n\,(2\pi)^D\delta(\sum\limits_{i=1}^n p_i)\,e^{-i\sum\limits_{i=1}^n p_i x_i}\,
e^{-\frac{i}{2}\!\!\!\!\!\!\sum\limits_{1\leq i < j< n}\!\!\!p_i\wedge p_j}}
\label{lackIR}\\
&&\times\,\frac{{\rm Tr}[(\slash k+\slashed p_1-m)\gamma^{\mu_1}(\slashed k-m)\gamma^{\mu_2}(\slashed k-\slashed p_2-m)\gamma^{\mu_3}\cdots
\gamma^{\mu_{n-1}}(\slashed k-\sum_{i=2}^{n-1}\slashed p_i-m)\gamma^{\mu_n}]}
{((k+p_1)^2+m^2)k^2((k-p_2)^2+m^2)\cdots ((k-\sum_{i=2}^{n-1}p_i)^2+m^2)}.
\nonumber
\end{eqnarray}
From (\ref{lackIR}) we draw the conclusion that the full one-loop $\Gamma_{Dirac}[b]$ in (\ref{Diracpath}) lacks any singular behaviour
when any of the momenta $\tilde{p}^\mu_i=\theta^{\mu\nu}\tilde{p}_{\nu\,i}$ vanishes $\forall i$.

Finally, it is plain that (\ref{oneloopfe}) and (\ref{lackIR}) hold whatever the gauge group and representation provided
the noncommutative Dirac field transforms as follows
\begin{equation}
\psi(b,\psi)\rightarrow i\Omega(b,\omega)\star\psi(b,\psi)
\label{4.15}
\end{equation}
under infinitesimal noncommutative gauge transformations defined by $\Omega(b,\omega)$, while $\omega(x)$ defines the infinitesimal ordinary gauge transformations.

\section{Conclusions and outlook}

The main conclusion of this paper is that noncommutative SU(N) defined by means of the $\theta$-exact
Seiberg-Witten map has a two-point function for the gauge field that exhibits UV/IR mixing at the one-loop,
in spite of the fact that there are no fundamental U(1) degrees of freedom. This is at odds with the noncommutative U(N) case where it is only the U(1) part of the one-loop two-point function  for the ordinary gauge field the one which is affected by the famous UV/IR mixing. Indeed, in the noncommutative U(N) case the famous noncommutative IR singularity in the one-loop two-point function of the ordinary gauge field reads \cite{VanRaamsdonk:2001jd}
\begin{equation}
{\frac{2g^2}{\pi^2}}\frac{\tilde{p}^\mu \tilde{p}^\nu}{(\tilde{p}^2)^2}\big[{\rm Tr}\,b^{\bot}_{\mu}(p)\big]\big[{\rm Tr}\, b^{\bot}_\nu(-p)\big].
\label{UNIRsingularity2}
\end{equation}
In Appendix C, we re-derive the above formula using the enveloping algebra approach together with the background field and the path integral methods employed in our recent works \cite{Martin:2016zon,Martin:2016hji,Martin:2016saw,Martin:2017nhg}, as well as in the previous sections of this paper, confirming eq. (\ref{UNIRsingularity2}) and showing this way implicitly the correctness of our computations. 

The noncommutative IR singularity of the noncommutative SU(N) theory reads 
\begin{equation}
\frac{g^2}{N}\frac{-2}{ 3\pi^2\tilde{p}^2}\Big\{\eta^{\mu\nu}+
2\frac{\tilde{p}^\mu \tilde{p}^\nu}{\tilde{p}^2}\Big\}{\rm Tr}\,\big[b^{\bot}_{\mu}(p) b^{\bot}_\nu(-p)\big],
\label{IRsingularity2}
\end{equation}
--see (\ref{IRsingularity})-- which has a novel, as yet unknown, UV/IR mixing tensor and Lie algebra structure with respect to that in (\ref{UNIRsingularity2}). Indeed, in (\ref{UNIRsingularity2}), we have the product of two traces over the generators of the Lie algebra so that only the U(1) part of the contribution gets affected by the noncommutative IR divergence.
  However, in (\ref{IRsingularity2}), we have only one trace over the product of two SU(N) generator and, hence, the full SU(N) part of the contribution receives the noncommutative IR divergence. No $\eta_{\mu\nu}$ occurs in (\ref{UNIRsingularity2}).

In subsection 3.4 -see (\ref{IR1}), (\ref{IR2}), (\ref{IR3}), etc-- we have shown that the noncommutative IR divergence in (\ref{IRsingularity2}) occurs because the noncommutative phase $e^{i\theta^{\mu\nu}k_{\mu}p_{\nu}}$ --$k$ being the loop momenta and $p$ the external momenta-- regularizes, provided $\theta^{\mu\nu}p_\nu\neq 0$, otherwise UV divergent integrals. Thus, the UV divergence is turned into an IR divergence at $\theta^{\mu\nu}p_\nu=0$. This the UV/IR mixing phenomenon and its origin has nothing to do with the use of the Seiberg-Witten map and its vanishing denominators at $\theta^{\mu\nu}k_{\mu}p_{\nu}=0$.

 We have also shown that this conclusion also holds in noncommutative QCD, since the addition of Dirac fermions in the fundamental representation to the noncommutative SU(N) theory does not modify the UV/IR mixing behaviour of the two-point function at hand.

The phenomenological implications of the UV/IR mixing unveiled in this paper are worth studying.
Notice that this UV/IR mixing  affects the noncommutative Standard Model of \cite{Calmet:2001na}
and the noncommuative GUTs models of \cite{Aschieri:2002mc, Martin:2013lba}
 when defined by means of the $\theta$-exact Seiberg-Witten map.

In the spirit of the Scattering amplitudes approach \cite{Raju:2009yx,Huang:2010fc,Arkani-Hamed:2017mur,Mizera:2019blq,Latas:2020nji} such a model has been recently successfully applied to estimate a  lower bound on the scale of noncommutativity \cite{Latas:2020nji,Horvat:2020ycy} by comparison with ATLAS-LHC  collaboration measurements of total cross sections for $\rm PbPb(\gamma\gamma)\to Pb^*Pb^*\gamma\gamma$ and $\rm PbPb(\gamma\gamma)\to Pb^*Pb^*\ell^+\ell^-$ reactions in the lead $^{208}$Pb ion-ion collision experiments \cite{Aaboud:2017bwk,Aad:2019ock}.


\appendix

\section{Seiberg-Witten maps}

Let $a_\mu$ and $c$ be, respectively, an ordinary gauge field and the corresponding ghost field taking values in a  Lie algebra in a given irreducible representation. A Seiberg-Witten map which  defines a noncommutative gauge field $A_{\mu}(a)$ and the corresponding noncommutative ghost field $C(a,c)$
in terms of ordinary $a_\mu$ and $c$ is obtained \cite{Barnich:2002tz, Cerchiai:2002ss, Barnich:2002pb} by solving, between $t=0$ and $t=1$,
 the following ``evolution'' problem
\begin{equation}
\begin{array}{l}
{\frac{d\phantom{t}}{dt}A_\mu(a;t)=-\frac{g}{4}\theta^{ij}\{A_i(a,t),\partial_j A_\mu(a,t)+F_{j\mu}(a,t)\}_{\star_t},\quad A_\mu(a;t=0)=a_\mu,}\\[8pt]
{\frac{d \phantom{t}}{dt}C(a,c;t)=\frac{g}{4}\theta^{ij}\{\partial_i C(a,c;t), A_j(a,t)\}_{\star_t},\quad C(a,c;t=0)=c.}
\label{SWevol}
\end{array}
\end{equation}
The product $\star_t$ has been defined in (\ref{tstar})  and
$F_{\mu\nu}(a;t)=\partial_\mu A_\nu(a;t)-\partial_\nu A_\mu(a;t)-ig[A_\mu(a;t),A_\nu(a;t)]_{\star_t}$.

The ``evolution'' problem in (\ref{SWevol}) can be solved \cite{Martin:2012aw}  by expanding in power of the coupling constant $g$ as befits the definition a field theory in perturbation theory:
\begin{equation}
\begin{array}{l}
{A_\mu(a)=a_\mu\,+\,g\, A_\mu^{(2)}(a,a;t=1)\,+\,g^2\, A_\mu^{(3)}(a,a,a;t=1)\,+\, O(g^3),}\\[8pt]
{C(a,c)=c\,+\,g\, C^{(2)}(a,c;t=1)+g^2\, C^{(3)}(a,c,a;t=1)\,+\, O(g^3),}
\label{theAS}
\end{array}
\end{equation}
with
\begin{equation}
\begin{array}{l}
\label{theASprime}
{A_\mu^{(2)}(a,a;t)=-\frac{1}{4}\theta^{ij}\int_0^t ds\,\{a_i,2\partial_j a_\mu-\partial_\mu a_j\}_{\star_s}},\\[10pt]
{A_\mu^{(3)}(a,a,a;t)=-\frac{1}{4}\theta^{ij}\int_0^t ds\Big[\{a_i,2\partial_j A^{(2)}_\mu(a,a;s)-2\partial_\mu A^{(2)}_j(a,a;s)\}_{\star_s}}\\[4pt]
{\phantom{A_\mu^{(3)}(a,a,a;t)=-\frac{1}{4}\theta^{ij}\int_0^t ds\Big[\{a_i,2}
+\{A^{(2)}_i(a,a;s),2\partial_j a_\mu-\partial_\mu a_j\}_{\star_s}-i\{a_i,[a_j,a_\mu]_{\star_s}\}_{\star_s}\Big]},\\[10pt]
{C^{(2)}(a,c;t)=\frac{1}{4}\theta^{ij}\int_0^t ds\,\{\partial_i c,\partial_i a_j\}_{\star_s}},\\[10pt]
{C^{(3)}(a,c,a;t)=\frac{1}{4}\theta^{ij}\int_0^t ds\,\Big[\{\partial_i C^{(2)}(a,c;s), a_j\}_{\star_s}+
\{\partial_i c, A^{(2)}_j(a,a;s) \}_{\star_s}\Big].}
\end{array}
\end{equation}

To obtain $B_\mu(b)$ and $Q_\mu(b,q)$ in (\ref{NCsplitting}), we replace $a_\mu$ with $b_\mu+q_\mu$ in (\ref{theAS}) and (\ref{theASprime}),
and then, expand in powers of $a_\mu$ and $b_\mu$ the resulting expressions:
\begin{equation}
\begin{array}{l}
{A_\mu(b+q)=B_\mu(b)+Q_\mu(b,q),}\\[10pt]
{B_\mu(b)=b_\mu+g\,\hat{A}^{(2)}_\mu(b,b)+g^2\, \hat{A}^{(3)}_\mu(b,b,b)+ O(g^3),}\\[8pt]
{\hat{A}^{(2)}_\mu(b,b)=A_\mu^{(2)}(a=b,a=b;t=1),\quad\hat{A}^{(3)}_\mu(b,b,b)=A_\mu^{(3)}(a=b,a=b,a=b;t=1),}\\[12pt]
{Q_\mu(b,q)=q_\mu+g\,\big(\hat{A}^{(2)}_\mu(b,q)+\hat{A}^{(2)}_\mu(q,q)\big)}\\[8pt]
{\qquad\qquad\qquad+g^2\,\big(\hat{A}^{(3)}_\mu(b,b,q)+\hat{A}^{(3)}_\mu(q,q,b)+\hat{A}^{(3)}_\mu(q,q,q)\big)+ O(g^3),}\\[4pt]
{\qquad\hat{A}^{(2)}_\mu(b,q)=-\frac{1}{4}\theta^{ij}\int_0^1 dt\,\Big[
\{b_i,2\partial_j q_\mu-\partial_\mu q_j\}_{\star_t}+\{q_i,2\partial_j b_\mu-\partial_\mu b_j\}_{\star_t}\Big],}\\[4pt]
{\qquad\hat{A}^{(2)}_\mu(q,q)= A_\mu^{(2)}(a=q,a=q;t=1),}\\[4pt]
{\qquad\hat{A}_\mu^{(3)}(q,q,b)=-\frac{1}{4}\theta^{ij}\int_0^1 dt}\\[8pt]
{\qquad\quad\phantom{Xx.}\times\Big[\{q_i,2\partial_j \hat{A}^{(2)}_\mu(b,q;t)-2\partial_\mu \hat{A}^{(2)}_j(b,q;t)\}_{\star_t}
+\{\hat{A}^{(2)}_i(b,q;t),2\partial_j q_\mu-\partial_\mu q_j\}_{\star_t}}\\[4pt]
{\phantom{\qquad\hat{A}_\mu^{(3)}xx.}+\{b_i,2\partial_j \hat{A}^{(2)}_\mu(q,q;t)-2\partial_\mu \hat{A}^{(2)}_j(b,q;t)\}_{\star_t}
+\{\hat{A}^{(2)}_i(q,q;t),2\partial_j b_\mu-\partial_\mu q_j\}_{\star_t}}\\[4pt]
{\phantom{\qquad\hat{A}_\mu^{(3)}xx}
-i\{q_i,[q_j,b_\mu]_{\star_t}\}_{\star_t}-i\{q_i,[b_j,q_\mu]_{\star_t}\}_{\star_t}-i\{b_i,[q_j,q_\mu]_{\star_t}\}_{\star_t}\Big],
}\\[4pt]
{\qquad\hat{A}_\mu^{(3)}(b,b,q)=\hat{A}_\mu^{(3)}(q\rightarrow b,q\rightarrow b,b\rightarrow q),}\\[4pt]
{\qquad\hat{A}_\mu^{(3)}(a,a,a)=A_\mu^{(3)}(a=q,a=q,a=q;t=1).}
\label{thehatAS}
\end{array}
\end{equation}

Analogously, $C(b+q,c)$ is obtained by setting $a_\mu=b_\mu+q_\mu$ in $C(a,c)$ in (\ref{theAS}):
\begin{equation}
\begin{array}{l}
C(b+q,c)= c + g\,\big(\hat{C}^{(2)}(c,b)+\hat{C}^{(2)}(q,c)\big)\\[10pt]
{\phantom{XXXXXXX}+g^2\,\big(\hat{C}^{(2)}(c,q,q)+\hat{C}^{(2)}(c,b,q)+\hat{C}^{(3)}(c,b,b)\big)+O(g^3)},\\[10pt]
{\hat{C}^{(2)}(b,c;t)=\frac{1}{4}\theta^{ij}\int_0^t ds\,\{\partial_i c, b_j\}_{\star_s}}\,,\\[10pt]
{\hat{C}^{(3)}(c,b,b;t)=\frac{1}{4}\theta^{ij}\int_0^t ds\,\Big[\{\partial_i \hat{C}^{(2)}(b,c;s), b_j\}_{\star_s}+
\{\partial_i c, \hat{A}^{(2)}_j(b,b;s) \}_{\star_s}\Big]}.
\end{array}
\label{A5}
\end{equation}
Since our purpose is to compute $\Gamma_2[b]$ at one-loop, we shall not need the $q$-dependent bits of $C(b+q,c)$:
\begin{equation}
C(b,c)= c + g\,\hat{C}^{(2)}(c,b)+g^2\,\hat{C}^{(3)}(c,b,b)+O(g^3),
\label{theCbc}
\end{equation}

Assume that $q_\mu$ and $b_\mu$ take values in the SU(N) Lie algebra in the fundamental representation. Then
$Q_\mu(b,q)$  in (\ref{thehatAS}) can be expressed as the following linear combination:
\begin{equation}
Q_\mu(b,q)= Q^{(0)}_\mu(b,q)\mathbb{I}_N\,+\, Q^{a}_\mu(b,q) T^a,
\label{theUNcombination}
\end{equation}
where $\mathbb{I}_N$ is the $N\times N$ identity matrix and $T^a$  are the SU(N) generators. We shall call  $Q^{(0)}_\mu(b,q)$ and
$Q^{a}_\mu(b,q)$ the U(1) component and SU(N) components of the noncommutative field $Q_\mu(b,q)$.

Let us introduce the following definitions:
\begin{equation}
\begin{array}{l}
{\hat{A}^{(2)\,0}_\mu(b,q)=\frac{1}{N}\,{\rm Tr}\,\hat{A}^{(2)}_\mu(b,q),\;\hat{A}^{(2)\,0}_\mu(q,q)=\frac{1}{N}\,{\rm Tr}\,\hat{A}^{(2)}_\mu(q,q),}\\[4pt]
{\hat{A}^{(3)\,0}_\mu(q,q,b)=\frac{1}{N}\,{\rm Tr}\,\hat{A}^{(3)}_\mu(q,q,b),
\;\hat{A}^{(3)\,0}_\mu(b,q,q)=\frac{1}{N}\,{\rm Tr}\,\hat{A}^{(2)}_\mu(b,q,q),}\\[4pt]
{\hat{A}^{(3)\,0}_\mu(q,q,q)=\frac{1}{N}\,{\rm Tr}\,\hat{A}^{(2)}_\mu(q,q,q),}\\[4pt]
{\hat{A}^{(2)\,a}_\mu(b,q)=\,{\rm Tr}\,(T^a\hat{A}^{(2)}_\mu(b,q)),\;\hat{A}^{(2)\,a}_\mu(q,q)=\,{\rm Tr}\,(T^a\hat{A}^{(2)}_\mu(q,q)),}\\[4pt]
{\hat{A}^{(3)\,a}_\mu(q,q,b)=\,{\rm Tr}\,(T^a\hat{A}^{(3)}_\mu(q,q,b)),
\;\hat{A}^{(3)\,a}_\mu(b,q,q)=\,{\rm Tr}\,(T^a\hat{A}^{(2)}_\mu(b,q,q)),}\\[4pt]
{\hat{A}^{(3)\,a}_\mu(q,q,q)=\,{\rm Tr}\,(T^a\hat{A}^{(2)}_\mu(q,q,q)).}
\label{thetraces}
\end{array}
\end{equation}
The reader should bear in mind the results in (\ref{thehatAS}). Then, taking into account (\ref{theUNcombination}), one  conclude that
\begin{equation}
\begin{array}{l}
{Q^{(0)}_\mu(b,q)=g\,\big(\hat{A}^{(2)\,0}_\mu(b,q)+\hat{A}^{(2)\,0}_\mu(q,q)\big)}\\[4pt]
{\phantom{XXXXx}+g^2\,\big(\hat{A}^{(3)\,0}_\mu(b,b,q)+\hat{A}^{(3)}_\mu(q,q,b)+\hat{A}^{(3)\,0}_\mu(q,q,q)\big)+ O(g^3),}\\[4pt]
{ Q^{a}_\mu(b,q)=
q^a_\mu+g\,\big(\hat{A}^{(2)\,a}_\mu(b,q)+\hat{A}^{(2)\,a}_\mu(q,q)\big)}\\[4pt]
{\phantom{XXXXXX,}+g^2\,\big(\hat{A}^{(3)\,a}_\mu(b,b,q)+\hat{A}^{(3)\,a}_\mu(q,q,b)+\hat{A}^{(3)\,a}_\mu(q,q,q)\big)+ O(g^3).}
\label{theQcomponents}
\end{array}
\end{equation}

Let us introduce the field $\tilde{q}_\mu=\tilde{q}^a_\mu T^a$, where
\begin{equation}
\begin{array}{l}
{\tilde{q}^a_\mu =Tr(T^a Q_\mu(b,q))=Q^{a}_\mu(b,q)=
q^a_\mu+g\,\big(\hat{A}^{(2)\,a}_\mu(b,q)+\hat{A}^{(2)\,a}_\mu(q,q)\big)+}\\[4pt]
{\phantom{\tilde{q}^a_\mu=}g^2\,\big(\hat{A}^{(3)\,a}_\mu(b,b,q)+\hat{A}^{(3)\,a}_\mu(q,q,b)+\hat{A}^{(3)\,a}_\mu(q,q,q)\big)+ O(g^3).}
\end{array}
\end{equation}
The previous expression can be inverted by expanding in powers of $g$:
\begin{equation}
q^a_\mu=\tilde{q}^a_\mu-g\,\big(\hat{A}^{(2)\,a}_\mu(b,\tilde{q})+\hat{A}^{(2)\,a}_\mu(\tilde{q},\tilde{q})\big)+O(g^2),
\label{theinversion}
\end{equation}
where
\begin{equation}
\hat{A}^{(2)\,a}_\mu(b,\tilde{q})=\hat{A}^{(2)\,a}_\mu(b,q=\tilde{q}),\;\hat{A}^{(2)\,a}_\mu(q=\tilde{q},q=\tilde{q}).
\label{A.12}
\end{equation}
Substituting (\ref{theinversion}) in (\ref{theUNcombination}) and then expanding in powers of $g$, one gets
\begin{equation}
\begin{array}{l}
{Q_\mu(b,q)=\tilde{q}_\mu+\mathbb{I}_N\Big\{g\big[\hat{A}^{(2)\,0}_\mu(b,\tilde{q})+
\hat{A}^{(2)\,0}_\mu(\tilde{q},\tilde{q})\big]}\\[4pt]
{\phantom{XXXXXXXX}+g^2\,\big[\hat{A}^{(3)\,0}_\mu(b,b,\tilde{q})+\hat{A}^{(3)\,a}_\mu(\tilde{q},\tilde{q},b)+
\hat{A}^{(3)\,a}_\mu(\tilde{q},\tilde{q},\tilde{q})}\\[4pt]
{\phantom{XXXXXXXXXx}-\hat{A}^{(2)\,0}_\mu(b,\hat{A}^{(2)\,a}_\sigma(b,\tilde{q})T^a)
-\hat{A}^{(2)\,0}_\mu(b,\hat{A}^{(2)\,a}_\sigma(\tilde{q},\tilde{q})T^a)}\\[4pt]
{\phantom{XXXXXXXXXx}-\hat{A}^{(2)\,0}_\mu(\tilde{q},\hat{A}^{(2)\,a}_\sigma(b,\tilde{q})T^a)
-\hat{A}^{(2)\,0}_\mu(\hat{A}^{(2)\,a}_\sigma(b,\tilde{q})T^a,\tilde{q})\big]\Big\},}
\label{theQbq}
\end{array}
\end{equation}
where
\begin{equation}
\begin{array}{l}
{\hat{A}^{(2)\,0}_\mu(b,\tilde{q})=\hat{A}^{(2)\,0}_\mu(b,q\rightarrow\tilde{q}),\;
\hat{A}^{(2)\,0}_\mu(\tilde{q},\tilde{q})=\hat{A}^{(2)\,0}_\mu(q=\tilde{q},q\rightarrow\tilde{q})},\\[4pt]
{\hat{A}^{(3)\,0}_\mu(b,b,\tilde{q})=\hat{A}^{(3)\,0}_\mu(b,b,q\rightarrow\tilde{q}),\;
\hat{A}^{(3)\,a}_\mu(\tilde{q},\tilde{q},b)=\hat{A}^{(3)\,a}_\mu(q\rightarrow\tilde{q},q\rightarrow\tilde{q},b),}\\[4pt]
{\hat{A}^{(3)\,0}_\mu(\tilde{q},\tilde{q},\tilde{q})=
\hat{A}^{(3)\,0}_\mu(q\rightarrow\tilde{q},q\rightarrow\tilde{q},q\rightarrow\tilde{q}),}\\[4pt]
{\hat{A}^{(2)\,0}_\mu(b,\hat{A}^{(2)\,a}_\sigma(b,\tilde{q})T^a)=
\hat{A}^{(2)\,0}_\mu(b,q_\sigma\rightarrow\hat{A}^{(2)\,a}_\sigma(b,\tilde{q})T^a),}\\[4pt]
{\hat{A}^{(2)\,0}_\mu(b,\hat{A}^{(2)\,a}_\sigma(\tilde{q},\tilde{q})T^a)=
\hat{A}^{(2)\,0}_\mu(b,q_\sigma\rightarrow\hat{A}^{(2)\,a}_\sigma(\tilde{q},\tilde{q})T^a),}\\[4pt]
{\hat{A}^{(2)\,0}_\mu(\hat{q},\hat{A}^{(2)\,a}_\sigma(b,\tilde{q})T^a)=-\frac{1}{4N}\theta^{ij}\int_0^1 dt\,{\rm Tr}\,
\big( \{\tilde{q}_i,2\partial_j \hat{A}^{(2)\,a}_\mu(b,\tilde{q})T^a-\partial_\mu \hat{A}^{(2)\,a}_j(b,\tilde{q})T^a\}_{\star_t}\big),}\\[4pt]
{\hat{A}^{(2)\,0}_\mu(\hat{q},\hat{A}^{(2)\,a}_\sigma(b,\tilde{q})T^a)=-\frac{1}{4N}\theta^{ij}\int_0^1 dt\,{\rm Tr}\,
\big( \{\hat{A}^{(2)\,a}_i(b,\tilde{q})T^a,2\partial_j \tilde{q}_\mu-\partial_\mu \tilde{q}_j\}_{\star_t}\big).}
\label{theqtildeAS}
\end{array}
\end{equation}
The notation $q_\mu\rightarrow object$ points out that $q_\mu$ is to be replaced with $object$ in the corresponding expression in (\ref{thetraces}) and (\ref{thehatAS}).

Proceeding in an analogous way, one shows that
\begin{equation}
C(b,c)=\tilde{c}+\mathbb{I}_N\big[g\,\hat{C}^{(2)\,0}(\tilde{c},b)+g^2\,\big(\hat{C}^{(3)\,0}(\tilde{c},b,b)+
\hat{C}^{(2)\,0}(\hat{C}^{(2)\,a}(\tilde{c},b)T^a,b)\big)\big]+O(g^3,b^3),
\label{thectilde}
\end{equation}
where
\begin{equation}
\begin{array}{l}
{\tilde{c}= \tilde{c}^a T^a,\;\tilde{c}^a=\,{\rm Tr}\,\big(T^a C(b,c)\big),}\\[4pt]
{\hat{C}^{(2)\,0}(\tilde{c},b)=\frac{1}{4N}\theta^{ij}\int_0^1 dt\,{\rm Tr}\,\big(\{\partial_i \tilde{c}, b_j\}_{\star_t}\big),}\\[4pt]
{C^{(3)\,0}(\tilde{c},b,b)=\frac{1}{4N}\theta^{ij}\int_0^1 dt\,{\rm Tr}\,\Big[\{\partial_i\hat{C}^{(2)}(\tilde{c},b;s), b_j\}_{\star_s}
+\{\partial_i \tilde{c}, \hat{A}^{(2)}_j(b,b;s) \}_{\star_s}\Big],}\\[4pt]
{\hat{C}^{(2)\,0}(\hat{C}^{(2)\,a}(\tilde{c},b)T^a,b)=\frac{1}{4N}\theta^{ij}\int_0^1 dt\,{\rm Tr}\,\big(\{\partial_i\hat{C}^{(2)\,a}(\tilde{c},b)T^a , b_j\}_{\star_t}\big),}\\[4pt]
{\hat{C}^{(2)\,a}(\tilde{c},b)=\frac{1}{4}\theta^{ij}\int_0^1 dt\,{\rm Tr}\,\big(T^a\{\partial_i \tilde{c}, b_j\}_{\star_t}\big).}
\end{array}
\label{A.16}
\end{equation}
Let us finally recall that (\ref{theQbq}) and  (\ref{thectilde}) are needed to go from (\ref{effectiveactionnew}) to (\ref{2pteffective}).

\section{Computing relevant integrals in $D$ dimensions for Euclidean signature }

Here for normalised $D$ dimensional integral we shall use the following shorthand notations $\int\frac{d^D k}{(2\pi)^D}\equiv\int$. Now we categorise integrals by the power of $(k\tilde p)$ factor in the denominator. First we have four integrals with power zero:

\begin{eqnarray}
I_1&=&\int\,\frac{e^{\pm i k\tilde p}}{k^2},\;\;
I_2=\int\,\frac{e^{i k\tilde p}}{k^2(k+p)^2},
\nonumber\\
I_3&=&\int\,\frac{(2k+p)^\mu(2k+p)^\nu}{k^2(k+p)^2}e^{i k\tilde p},\;\;
I_4=\int\,\frac{(2k+p)^\mu(2k+p)^\nu}{k^2(k+p)^2}.
\label{B1}
\end{eqnarray}
Then we have six with power one:

\begin{equation}
I_5=\int\,\frac{1}{k^2(k\tilde p)},\;\;
I_6=\int\, \frac{k^\mu}{k^2(k\tilde p)},\;\;
I_7=\int\, \frac{k^\mu k^\nu}{k^2(k\tilde p)},
\label{B2}
\end{equation}
\begin{equation}
I_8=\int\,\frac{e^{i k\tilde p}+e^{-i k\tilde p}}{k^2(k\tilde p)},\;\;
I_9=\int\,\frac{k^\mu\big(e^{i k\tilde p}+e^{-i k\tilde p}\big)}{k^2(k\tilde p)},\;\;
I_{10}=\int\,\frac{k^\mu k^\nu\big(e^{i k\tilde p}+e^{-i k\tilde p}\big)}{k^2(k\tilde p)},
\label{B3}
\end{equation}
and finally comes seven with power two:
\begin{equation}
I_{11}=\int\,\frac{1}{k^2(k\tilde p)^2},\;\;
I_{12}=\int\,\frac{k^\mu}{k^2(k\tilde p)^2},\;\,
I_{13}=\int\,\frac{k^\mu k^\nu}{k^2(k\tilde p)^2},
\label{B4}
\end{equation}
\begin{equation}
I_{14}=\int\,\frac{1}{k^2(k\tilde p)^2}\big(e^{i k\tilde p}+e^{-i k\tilde p}\big),\;\;
I_{15}=\int\,\frac{k^\mu}{k^2(k\tilde p)^2}\big(e^{i k\tilde p}+e^{-i k\tilde p}\big),
\label{B5}
\end{equation}
\begin{equation}
I_{16}=\int\,\frac{k^\mu k^\nu}{k^2(k\tilde p)^2}\big(e^{i k\tilde p}+e^{-i k\tilde p}\big),\;\;
I_{17}=\int\frac{k^\mu k^\nu k^\rho}{k^2(k\tilde p)^2}\big(e^{i k\tilde p}+e^{-i k\tilde p}\big).
\label{B6}
\end{equation}
Using $k\to -k$ trick we can find immediately that:
$I_5=I_7=I_{10}=I_{12}=I_{15}=I_{17}=0$.
We can also transform $k\to-k-p$ and show that
\begin{equation}
I_3=\int\,\frac{2k^\mu k^\nu+k^\mu p^\nu}{k^2(k+p)^2}(e^{i k\tilde p}+e^{-i k\tilde p}),\;\;
I_4=2\int\,\frac{2k^\mu k^\nu+k^\mu p^\nu}{k^2(k+p)^2}.
\label{B8}
\end{equation}
From history we know the $I_{1,2}$ integrals:
\begin{equation}
I_1=\frac{1}{(4\pi)^{\frac{D}{2}}}\left(\frac{\tilde p^2}{4}\right)^{1-\frac{D}{2}}\Gamma\left(\frac{D}{2}-1\right),
\label{B9}
\end{equation}
\begin{equation}
I_2=\frac{2}{(4\pi)^{\frac{D}{2}}}\int\limits_0^1 dx \big(x(1-x)p^2\big)^{\frac{D}{4}-1}\left(\frac{\tilde p^2}{4}\right)^{1-\frac{D}{4}}K_{\frac{D}{2}-2}
\left[\sqrt{x(1-x)p^2\tilde p^2}\right].
\label{B10}
\end{equation}
Integral $I_3$ is a standard nonplanar integral, on which we apply the decomposition
\begin{equation}
I_3= \eta^{\mu\nu} p^2\cdot I_{3_1}+(p^\mu p^\nu -\eta^{\mu\nu} p^2)\cdot I_{3_2}+\tilde p^\mu \tilde p^\nu\cdot I_{3_3},
\label{B11}
\end{equation}
and after integration over parameter $\alpha$ obtain
\begin{gather}
I_{3_1}=\frac{2}{\tilde p^2} I_1,
\nonumber\\
I_{3_2}=\frac{2}{(4\pi)^{\frac{D}{2}}}\int\limits_0^1 dx (1-2x)^2\big(x(1-x)p^2\big)^{\frac{D}{4}-1}\left(\frac{\tilde p^2}{4}\right)^{1-\frac{D}{4}}K_{\frac{D}{2}-2}\left[\sqrt{x(1-x)p^2\tilde p^2}\right],
\nonumber\\
I_{3_3}=-\frac{2}{(4\pi)^{\frac{D}{2}}}\int\limits_0^1 dx \big(x(1-x)p^2\big)^{\frac{D}{4}}\left(\frac{\tilde p^2}{4}\right)^{-\frac{D}{4}}K_{\frac{D}{2}}\left[\sqrt{x(1-x)p^2\tilde p^2}\right].
\label{B12}
\end{gather}
Here $I_{3_3}$ can be re-expressed via $I_{3_1}$ and $K_{\frac{D}{2}-2}\left[\sqrt{x(1-x)p^2\tilde p^2}\right]$ as follows
\begin{equation}
\begin{split}
&I_{3_3}=(2-D)
I_{3_1}
\label{B13}\\&
+\frac{2}{(4\pi)^{\frac{D}{2}}}\frac{p^2}{\tilde p^2}
\int\limits_0^1 dx \big(4(D-1)x^2-D\big)\big(x(1-x)p^2\big)^{\frac{D}{4}-1}\left(\frac{\tilde p^2}{4}\right)^{1-\frac{D}{4}}K_{\frac{D}{2}-2}\left[\sqrt{x(1-x)p^2\tilde p^2}\right].
\end{split}
\end{equation}
Integral $I_4$ is a standard planar integral and we have the standard tensor reduction result:
\begin{equation}
\begin{split}
I_4=
\frac{1}{(4\pi)^{\frac{D}{2}}}\big(\eta^{\mu\nu}p^2 - p^\mu p^\nu\big)(p^2)^{\frac{D}{2}-2}B\left(\frac{D}{2}-1,\frac{D}{2}-1\right)\frac{\Gamma\left(2-\frac{D}{2}\right)}{1-D}.
\end{split}
\label{B14}
\end{equation}
Trivially one can see that
$I_6=I_{13}\;\tilde p_\nu,\;\, I_9=I_{16}\;\tilde p_\nu$, and
integrals $I_{13}$ and $I_{16}$ can be solved by an NC tensor reduction. Since the only relevant momentum
in these two integrals is $\tilde p^\mu$, we can prescribe the following simple tensor structures:
\begin{gather}
I_{13}=\eta^{\mu\nu}\tilde p^2A_{13}+\tilde p^\mu \tilde p^\nu B_{13},
\nonumber\\
I_{16}=\eta^{\mu\nu}\tilde p^2A_{16}+\tilde p^\mu \tilde p^\nu B_{16},
\label{B15}
\end{gather}
which after contraction with $\eta_{\mu\nu}$ and $\tilde p_\mu \tilde p_\nu$ gives: \begin{equation}
I_{13}=0=I_6,\;\,
I_{16}=\frac{2}{ \tilde p^4} \frac{\eta^{\mu\nu}\tilde p^2-\tilde p^\mu \tilde p^\nu\cdot D}{1-D}I_1 .
\label{B16}
\end{equation}
We are now left with $I_{11}$ and $I_{14}$, and they are not as easy as they seem to be.
Priorly we evaluated them in \cite{Martin:2016zon} as the following sum denoted as $T_{-2}$:
\begin{equation}
T_{-2}=2I_{11}-I_{14}.
\label{B17}
\end{equation}
Two methods were used to calculate $T_{-2}$ explicitly, one is based on Grozin parametrization \cite{Grozin:2000cm} and the other contour integral parametrization.
We encounter immediately problem with the second method because the poles at $x=0$ are second order in $I_{11}$ and $I_{14}$ while first order in $T_{-2}$,
which means that we can not define the principle value around this pole for $I_{11}$ or $I_{14}$.
This leaves only the Grozin parametrization producing:
\begin{equation}
\begin{split}
I_{11}=&-\int\limits_0^\infty dy\, y\int\limits_0^\infty d\alpha\, \alpha^2 e^{-\alpha (k^2+i y k\tilde p)}
\label{B18}
=\frac{-1}{(4\pi)^{\frac{D}{2}}}
\left(\frac{\tilde p^2}{4}\right)^{\frac{D}{2}-3}\Gamma\left(3-\frac{D}{2}\right)B\left(D-4, 4-D\right)=0,
\end{split}
\end{equation}
\begin{equation}
\begin{split}
I_{14}=
-2\int\limits_0^\infty dy\, y\int\limits_0^\infty d\alpha\, \alpha^2\int e^{-\alpha (k^2+i y k\tilde p)+i k\tilde p}
=
\frac{1}{(4\pi)^{\frac{D}{2}}}\left(\frac{\tilde p^2}{4}\right)^{1-\frac{D}{2}}\frac{\Gamma\left(\frac{D}{2}-2\right)}{3-D}.\end{split}
\end{equation}

\subsection{Discussing the UV divergent tadpole integral $I_{14}$}

In our recent computation a new type of tadpole integral,
which is UV divergent at the $D\to 4-\epsilon$ limit occurred repeatedly.
Here we provide an account of its evaluation.
This new tadpole $I_{14}$, in \cite{Martin:2016zon} denoted as $T_{-2}$,
bears a very simple form
\begin{equation}
I_{14}\equiv-T_{-2}=\int\frac{d^D \ell}{(2\pi)^D}\,\frac{f_{\star_2}(\ell,p)^2}{\ell^2}.
\label{B}
\end{equation}
On the other hand, it turns out that $T_{-2}$ is not quite easy to evaluate. Two usual regularization methods used before,
turning tadpole to bubble or using the $n$-nested zero regulator respectively, did not function here. The first one produces divergent special function integrals
while the second contains unfavourable powers of the regulator. The parametrization discussed in the first section of this note offers
us an alternative way to handle this problem, using this parametrization we can express $T_{-2}$ as
\begin{equation}
\begin{split}
T_{-2}&=\int\frac{d^{D-1} \ell}{(2\pi)^{D-1}}\,\int\limits_{-\infty}^{+\infty}\,\frac{dx}{2\pi}\frac{1}{\ell^2+x^2}\frac{4\sin^2\frac{|\tilde p|}{2}x}{x^2\tilde p^2}
=\frac{1}{\tilde p^2}\int\frac{d^{D-1} \ell}{(2\pi)^{D-1}}\,\left(-\frac{1}{|\ell|^3}+\frac{2|\tilde p|}{|\ell|^2}+\frac{e^{-|\ell||\tilde p|}}{|\ell|^3}\right).
\end{split}
\label{B}
\end{equation}
Here we can only neglect the second term in the last parenthesis because the first and last exceed the minimal power
of $|\ell|$ for massless tadpole to vanish in the dimensional regularization prescription. One can introduce one more integrand $y$ to make
he first plus the last terms into one:
\begin{equation}
\begin{split}
T_{-2}&=\frac{1}{\tilde p^2}\int\frac{d^{D-1} \ell}{(2\pi)^{D-1}}\,\left(-\frac{1}{|\ell|^3}+\frac{e^{-|\ell||\tilde p|}}{|\ell|^3}\right)
=\frac{1}{(4\pi)^{\frac{D}{2}}}\left(\frac{\tilde p^2}{4}\right)^{1-\frac{D}{2}}\frac{\Gamma\left(\frac{D}{2}-2\right)}{D-3}.
\end{split}
\label{B}
\end{equation}
A familiar pattern emerges once we compute the $D\to 4+2\epsilon$ limit
\begin{equation}
T_{-2}=\frac{-4}{(4\pi)^2\tilde p^2}\left(-\frac{1}{\epsilon}+Ln\,\tilde p^2+Ln(\pi\mu^2)+\Gamma_E+2\right)+\mathcal O(\epsilon).
\label{B}
\end{equation}
Here we see the logarithmic UV/IR mixing taking place via a single integral.

\subsection{The $D\to 4+2\epsilon$ limit of all evaluated integrals}
\begin{eqnarray}
I_1&=&\frac{1}{4\pi^2}\frac{1}{\tilde p^2}+\mathcal O(\epsilon),
\label{I_1}\\
I_2&=&\frac{1}{8\pi^2}\int\limits_0^1 dx\, K_0\left[\sqrt{x(1-x)p^2\tilde p^2}\right]+\mathcal O(\epsilon),
\label{I_2}\\
I_3&=& \frac{1}{8\pi^2}\Big(\eta^{\mu\nu} \frac{4}{\tilde p^2}
+\big(p^\mu p^\nu -\eta^{\mu\nu} p^2\big)\int\limits_0^1 dx (1-2x)^2K_0\left[\sqrt{x(1-x)p^2\tilde p^2}\right]\Big)
\nonumber\\&-&\frac{1}{\pi^2}\frac{\tilde p^\mu \tilde p^\nu}{\tilde p^4}
+\frac{1}{2\pi^2}\frac{\tilde p^\mu \tilde p^\nu}{\tilde p^2}\int\limits_0^1 dx (3x^2-1)p^2K_0\left[\sqrt{x(1-x)p^2\tilde p^2}\right]
+\mathcal O(\epsilon),
\label{I_3}\\
I_4&=&\frac{1}{48\pi^2}\big(\eta^{\mu\nu}p^2 - p^\mu p^\nu\big)\left(\frac{1}{\epsilon}+\Gamma_E+Ln\frac{p^2}{4\pi\mu^2}-\frac{8}{3}\right)+\mathcal O(\epsilon),
\label{I_4}\\
I_5&=&I_6=I_7=I_8=I_{10}=I_{11}=I_{12}=I_{13}=I_{15}=I_{17}=0,
\label{I=0}\\
I_9&=&\frac{1}{2\pi^2}\frac{\tilde p^\mu}{\tilde p^4}+\mathcal O(\epsilon),
\label{I_9}\\
I_{14}&=&-T_{-2}=\frac{1}{(2\pi)^2}\frac{1}{\tilde p^2}\left(-\frac{1}{\epsilon}+Ln\,\tilde p^2+Ln(\pi\mu^2)+\Gamma_E+2\right)+\mathcal O(\epsilon),
\label{I_14}\\
I_{16}&=&-\frac{1}{6\pi^2}\frac{1}{\tilde p^6}\big(\eta^{\mu\nu}\tilde p^2-4\tilde p^\mu \tilde p^\nu\big)+\mathcal O(\epsilon).
\label{I_16}
\end{eqnarray}

\section{Rederiving the noncommutative IR divergence for the two-point function of U(N) in the fundamental representation.}

In this Appendix we shall apply the techniques displayed in Sections 2 and 3 to the U(N) case in the fundamental representation and thus rederive the result in eq. (\ref{UNIRsingularity2}). Actually, the techniques in Sections 2 and 3 have been previously applied in ref. \cite{Martin:2016saw} to show perturbatively the duality of noncommutative U(N) gauge theory under the Seiberg-Witten map.

Let us assume that $a_\mu$, $b_\mu$ and $q_\mu$ in Section 2 --see eq. (\ref{splitting}), in particular-- take values in the Lie algebra of U(N) in the fundamental representation. Then, $A_\mu(b+q)$, $B_\mu(b)$ and $Q_\mu(b,q)$ --see eq. (\ref{NCsplitting})-- as defined by the Seiberg-Witten map in Appendix A also take values in the Lie algebra of U(N) in the fundamental representation --notice that this not so for SU(N). Analogously, let us assume that $c$, $\Bar{C}$, $F$ in section 2 --see eq. (\ref{ordinaryBRS}), in particular-- are elements of the Lie algebra of U(N) in the fundamental representation. Then, $C(b+q,c)$ as defined by the Seiberg-Witten map in Appendix A is an element of U(N) in the fundamental representation. It is not difficult to convince oneself  one can proceed as in Sections 2 and 3 --by adapting the formulae in those Sections to the U(N) case at hand and with the help of the results in ref. \cite{Martin:2016saw}-- and compute the off-shell two-point contribution to effective action --let us call it $\Gamma^{\rm U(N)}_{ 2}[b]$-- of the DeWitt effective action. 
The $\Gamma^{\rm U(N)}_{ 2}[b]$ reads
\begin{equation}
\begin{array}{l}
{\Gamma^{\rm U(N)}_2[b]=g^2\int\frac{d^4 p}{(2\pi)^4} \,[\rm Tr\,b_{\mu}(p)][{\rm Tr}\, b_\nu(-p)]
\frac{2}{\pi^2}\frac{\tilde{p}^\mu \tilde{p}^\nu}{(\tilde{p}^2)^2}}\\[3pt]
{\phantom{\Gamma^{(U(N))}_2[b]}+\frac{g^2}{16 \pi^2}\int\frac{d^4 p}{(2\pi)^4} \,[{\rm Tr}\,b_{\mu}(p)][{\rm Tr}\,b_\nu(-p)]\big[(p^2\eta^{\mu\nu}-p^\mu p^\nu)\,
\frac{11}{3}\,Ln (p^2\tilde{p}^2)\,+\,\Sigma_{\mu\nu}(p,\tilde{p})\big]}
\\[3pt]
{\phantom{\Gamma^{(U(N))}_2[b]}+\frac{g^2}{16 \pi^2}N\int\frac{d^4 p}{(2\pi)^4} \,
{\rm Tr}\,\big[b_{\mu}(p)\big(p^2\eta^{\mu\nu}-p^\mu p^\nu\big)b_\nu(-p)\big]}\\[3pt]
{\phantom{\Gamma^{(U(N))}_2[b]+\frac{g^2}{16 \pi^2}N\int\frac{d^4 p}{(2\pi)^4}}\times \Big\{-\frac{11}{3}\Big(\frac{1}{\epsilon}+ Ln\frac{p^2}{4\pi\mu^2}\Big)+\frac{67}{144\pi^2}+ {\cal O}(D-4)\Big\}}\\[3pt]
{\phantom{\Gamma^{\rm(U(N))}_2[b]}+\;\Gamma^{\rm U(N)}_2[b]^{\rm (eom0)}+\;({\rm 2-loop\;order}),}
\end{array}
\label{uncase}
\end{equation}
where $\Gamma^{\rm U(N)}_2[b]^{\rm(eom0)}$, which obviously vanishes upon imposing the equation of motion, contains all the contributions on which integral
\begin{equation*}
\int d^4 x\,q^a_{\mu}(x)\frac{\delta \Gamma^{\rm U(N)}[b]}{\delta b^a_{\mu}(x)},
\end{equation*}
in (\ref{effectiveaction}), is involved. The $\Gamma^{\rm U(N)}_2[b]^{\rm(eom0)}$ is not physically relevant for it vanishes on-shell, while $\Sigma_{\mu\nu}(p,\tilde{p})$ is a function such that it remains finite as $\tilde{p}$ goes to zero.

It is apparent that if, in $\Gamma^{\rm U(N)}_2[b]$ above, we  replace $b_\mu$ with $b^{\bot}_\mu$ taking values in the Lie algebra of U(N) in the fundamental representation and satisfying the eqs. in (\ref{transverse}), we will obtain  eq. (\ref{UNIRsingularity2}) as expected. Hence, in the U(N) in the fundamental representation the color structure of the noncommutative IR divergence of the one-loop two-point function is $[{\rm Tr}\,b^{\bot}_\mu(p)][{\rm Tr}\,b^{\bot}_\mu(-p)]$, i.e, it is purely U(1), whereas in the SU(N) case is quite different: ${\rm Tr}\,[b^{\bot}_\mu(p) b^{\bot}_\mu(-p)]$, 
i.e. there is only one ${\rm Tr}$.

The $\Gamma^{\rm U(N)}_2[b]$ in eq. (\ref{uncase}) is the counterpart of $\Gamma_2[b]$ in eq. (\ref{quiteacomputation}). The fact that the result in eq. (\ref{quiteacomputation}) has a notably different tensor structure from that in eq. (\ref{uncase}) has to do with the fact in the U(N) case the change of integration variables in eq. (\ref{3.5}) removes from the classical action and the gauge-fixing terms all the $\theta$-dependence coming from the Seiberg-Witten map for $Q_\mu(b,q)$ and $C(b,q)$. This is not so in the SU(N) case, for in this case neither $Q_\mu(b,q)$ nor $C(b,q)$ belong to the Lie algebra of SU(N). On the other hand, both $Q_\mu(b,q)$ and $C(b,q)$ belong to the Lie algebra of U(N) in the fundamental representation if $b_\mu$, $q_\mu$ and $c$ take values in the Lie algebra of U(N) in the fundamental representation. The gauge dependent contributions in $\Gamma^{\rm U(N)}_2[b]$ and $\Gamma_2[b]$ are removed by going on-shell as required by DeWitt's effective action formalism. This procedure yields eqs. (\ref{UNIRsingularity2}) and (\ref{IRsingularity2}) for U(N) and SU(N), respectively, with tensor structures and --color structures-- quite different.

\acknowledgements{
The work by C.P. Martin has been financially supported in part by the Spanish MINECO through grant PGC2018-095382-B-I00. J.Trampetic would like thank Dieter L\"ust for many discussions and to acknowledge support of Max-Planck-Institute for Physics, M\"unchen. The work of J.You has been supported by Croatian Science Foundation.}



\begin{thebibliography}{999vs}

\bibitem{Kontsevich:1997vb}
  M.~Kontsevich,
 {\it Deformation quantization of Poisson manifolds. 1.}
  Lett.\ Math.\ Phys.\  {\bf 66} (2003) 157,  doi:10.1023/B:MATH.0000027508.00421.bf
  [q-alg/9709040].

\bibitem{Madore:2000en}
  J.~Madore, S.~Schraml, P.~Schupp and J.~Wess,
  {\it Gauge theory on noncommutative spaces,}
  Eur.\ Phys.\ J.\ C {\bf 16} (2000) 161,  doi:10.1007/s100520050012
  [hep-th/0001203].

\bibitem{Jurco:2000fb}
  B.~Jurco and P.~Schupp,
 {\it Noncommutative Yang-Mills from equivalence of star products,}
  Eur.\ Phys.\ J.\ C {\bf 14} (2000) 367,  doi:10.1007/s100520000380
  [hep-th/0001032].

\bibitem{Jurco:2001rq}
  B.~Jurco, L.~Moller, S.~Schraml, P.~Schupp and J.~Wess,
  {\it Construction of nonAbelian gauge theories on noncommutative spaces,}
  Eur.\ Phys.\ J.\ C {\bf 21} (2001) 383,  doi:10.1007/s100520100731
  [hep-th/0104153].

  \bibitem{Jurco:2001kp}
  B.~Jurco, P.~Schupp and J.~Wess,
 {\it Noncommutative line bundle and Morita equivalence,}
  Lett.\ Math.\ Phys.\  {\bf 61}, 171 (2002) [hep-th/0106110].


\bibitem{arXiv0711.2965B}
Martin Bordemann, Nikolai Neumaier, Stefan Waldmann, Stefan Weiß,
{\it Deformation quantization of surjective submersions and principal fibre bundles}. Crelle's J. reine angew. Math. 639 (2010), 1--38 [Abstract] [PDF] [MR2608189] [Zbl05687061], arXiv:0711.2965.

\bibitem{arXiv0909.4259B}
Henrique Burzstyn, Vasiliy Dolgushev, Stefan Waldmann,
{\it Morita equivalence and characteristic classes of star products.} Crelle's J. reine angew. Math. 662 (2012), 95-163. [Abstract] [PDF] [MR2876262] [Zbl1237.53080], arXiv:0909.4259.

  \bibitem{Filk:1996dm}
T.~Filk,
{\it {Divergencies in a field theory on quantum space}},
Phys. Lett. {\bf B376} (1996) 5, doi:10.1016/0370-2693(96)00024-X.

\bibitem{Martin:1999aq}
C.P.~Martin, D.~Sanchez-Ruiz,
{\it The One-loop UV Divergent Structure of {\rm U(1)} Yang-Mills Theory on Noncommutative $R^4$},
Phys. Rev. Lett.  {\bf 83} (1999) 476--479, [hep-th/9903077].

  \bibitem{Seiberg:1999vs}
N.~Seiberg and E.~Witten,
{\it String theory and noncommutative geometry},
  JHEP {\bf 09} (1999) 032,
  [arXiv:hep-th/9908142].

\bibitem{Okawa:2001mv}
  Y.~Okawa and H.~Ooguri,
{\it An Exact solution to Seiberg-Witten equation of noncommutative gauge theory,}
  Phys.\ Rev.\ D {\bf 64} (2001) 046009,  [hep-th/0104036].

\bibitem{Brace:2001fj}
  D.~Brace, B.~L.~Cerchiai, A.~F.~Pasqua, U.~Varadarajan and B.~Zumino,
  {\it A Cohomological approach to the nonAbelian Seiberg-Witten map,}
  JHEP {\bf 0106}, 047 (2001),  [hep-th/0105192].

\bibitem{Barnich:2002tz}
  G.~Barnich, F.~Brandt and M.~Grigoriev,
  {\it Seiberg-Witten maps in the context of the antifield formalism,}
  Fortsch.\ Phys.\  {\bf 50} (2002) 825
  doi:10.1002/1521-3978(200209)50:8/9<825::AID-PROP825>3.0.CO;2-V
  [hep-th/0201139].

\bibitem{Cerchiai:2002ss}
  B.~L.~Cerchiai, A.~F.~Pasqua and B.~Zumino,
  {\it The Seiberg-Witten map for noncommutative gauge theories,} in Continuous Advances in QCD 2002 / ARKADYFEST (honoring the 60th birthday of Prof. Arkady Vainshtein), 
pages 207-420, hep-th/0206231.

\bibitem{Barnich:2002pb}
  G.~Barnich, F.~Brandt and M.~Grigoriev,
  {\it Seiberg-Witten maps and noncommutative Yang-Mills theories for arbitrary gauge groups,}
  JHEP {\bf 0208} (2002) 023
  doi:10.1088/1126-6708/2002/08/023
  [hep-th/0206003].

\bibitem{Banerjee:2004rs}
R.~Banerjee and H.~S.~Yang,
{\it Exact Seiberg-Witten map, induced gravity and topological invariants in noncommutative field theories,} Nucl. Phys. B \textbf{708} (2005), 434-450, doi:10.1016/j.nuclphysb.2004.12.003
[arXiv:hep-th/0404064 [hep-th]].

\bibitem{Martin:2012aw}
  C.~P.~Martin,
  {\it Computing the $\theta$-exact Seiberg-Witten map for arbitrary gauge groups,}
  Phys.\ Rev.\ D {\bf 86} (2012) 065010
  doi:10.1103/PhysRevD.86.065010
  [arXiv:1206.2814 [hep-th]].

 \bibitem{Horvat:2011qn}
R.~Horvat, A.~Ilakovac, P.~Schupp, J.~Trampeti\'{c}, and J.~You,
{\it Yukawa couplings and seesaw neutrino masses in noncommutative gauge theory},
 {\em Phys. Lett.} {\bf B715}, 340-347 (2012),

\bibitem{Trampetic:2015zma}
  J.~Trampetic and J.~You,
  {\it $\theta$-exact Seiberg-Witten maps at order $e^3$,}
  Phys.\ Rev.\ D {\bf 91} (2015) no.12,  125027,  doi:10.1103/PhysRevD.91.125027,  [arXiv:1501.00276 [hep-th]].

  \bibitem{Bigatti:1999iz}
  Bigatti D., Susskind L.,
{\it  Magnetic fields, branes and noncommutative geometry},
Phys. Rev. D \textbf{62} (2000), 066004, doi:10.1103/PhysRevD.62.066004 [arXiv:hep-th/9908056 [hep-th]].

\bibitem{Minwalla:1999px}
  S.~Minwalla, M.~Van Raamsdonk and N.~Seiberg,
 {\it Noncommutative perturbative dynamics,}
  JHEP {\bf 0002}, 020 (2000),  [arXiv:hep-th/9912072].

\bibitem{Hayakawa:1999yt}
M.~Hayakawa,
{\it Perturbative analysis on infrared aspects of noncommutative QED on  R**4,}
Phys.\ Lett.\  B {\bf 478}, 394 (2000),  [arXiv:hep-th/9912094].

 \bibitem{Hayakawa:1999zf}
M.~Hayakawa,
{\it {Perturbative analysis on infrared and ultraviolet aspects of  noncommutative QED on $R^4$}},
[{{\tt hep-th/9912167}}].

\bibitem{Matusis:2000jf}
A.~Matusis, L.~Susskind, and N.~Toumbas,
{\it The IR/UV connection in the  non-commutative gauge theories},
 {\em JHEP} {\bf 12} (2000) 002,
 [arXiv:hep-th/0002075].

\bibitem{VanRaamsdonk:2000rr}
  M.~Van Raamsdonk and N.~Seiberg,
{\it Comments on noncommutative perturbative dynamics,} JHEP {\bf 0003} (2000) 035, doi:10.1088/1126-6708/2000/03/035, [hep-th/0002186].

\bibitem{VanRaamsdonk:2001jd}
M.~Van Raamsdonk,
{\it The Meaning of infrared singularities in noncommutative gauge theories,'}
JHEP \textbf{11} (2001), 006, doi:10.1088/1126-6708/2001/11/006 [arXiv:hep-th/0110093].

\bibitem{Hayakawa:2000zi}
  M.~Hayakawa,
{\it Perturbative ultraviolet and infrared dynamics of noncommutative  quantum
 field theory,}  in 30th International Conference on High-Energy Physics, pages 1455-1460, 
hep-th/0009098.
  
\bibitem{Ruiz:2000hu}
F.~R.~Ruiz,
{\it Gauge fixing independence of IR divergences in noncommutative U(1), perturbative tachyonic instabilities and supersymmetry,} Phys. Lett. B \textbf{502} (2001), 274-278, doi:10.1016/S0370-2693(01)00145-9
[arXiv:hep-th/0012171 [hep-th]].

\bibitem{Khoze:2000sy}
V.~V.~Khoze and G.~Travaglini,
{\it Wilsonian effective actions and the IR / UV mixing in noncommutative gauge theories,}
JHEP \textbf{01} (2001), 026, doi:10.1088/1126-6708/2001/01/026,[arXiv:hep-th/0011218 [hep-th]].

\bibitem{Armoni:2003va}
  A.~Armoni, E.~Lopez and A.~M.~Uranga,
 {\it Closed strings tachyons and noncommutative instabilities,} JHEP {\bf 0302} (2003) 020, doi:10.1088/1126-6708/2003/02/020, [hep-th/0301099].

\bibitem{Ferrari:2004ex}
A.~F.~Ferrari, H.~O.~Girotti, M.~Gomes, A.~Y.~Petrov, A.~A.~Ribeiro, V.~O.~Rivelles and A.~J.~da Silva,
{\it Towards a consistent noncommutative supersymmetric Yang-Mills theory: Superfield covariant analysis,}
Phys. Rev. D \textbf{70} (2004), 085012, doi:10.1103/PhysRevD.70.085012m [arXiv:hep-th/0407040 [hep-th]].

    \bibitem{Zeiner:2007}
J.~Zeiner, {\it Noncommutative quantumelectrodynamics from Seiberg-Witten Maps to all orders in Theta(mu nu)} (Wurzburg U.). Jul 2007. 139 pp.
\newblock PhD thesis.

    \bibitem{Schupp:2008fs}
P.~Schupp and J.~You,
{\it {UV/IR mixing in noncommutative QED defined by  Seiberg-Witten map}},
  JHEP {\bf 08} (2008) 107,
 [arXiv:hep-th/0807.4886].


\bibitem{Horvat:2011bs}
  R.~Horvat, A.~Ilakovac, J.~Trampetic and J.~You,
{\it On UV/IR mixing in noncommutative gauge field theories,}
JHEP {\bf 12} (2011) 081, arXiv:1109.2485 [hep-th].

\bibitem{Horvat:2011qg}
  R.~Horvat, A.~Ilakovac, P.~Schupp, J.~Trampetic and J.~You,
 {\it Neutrino propagation in noncommutative spacetimes,}
  JHEP {\bf 1204} (2012) 108,  [arXiv:1111.4951 [hep-th]].

\bibitem{Horvat:2013rga}
  Horvat R., Ilakovac A., Trampetic J., You J.,
 Self-energies on deformed spacetimes,
 JHEP {\bf 1311} (2013) 071,  [arXiv:1306.1239].

\bibitem{Horvat:2015aca}
  R.~Horvat, J.~Trampetic and J.~You,
 {\it Photon self-interaction on deformed spacetime,}
  Phys.\ Rev.\ D {\bf 92} (2015) no.12,  125006,  doi:10.1103/PhysRevD.92.125006,  [arXiv:1510.08691 [hep-th]].


















\bibitem{Grosse:2005iz}
  H.~Grosse and M.~Wohlgenannt,
  {\it On $\kappa$-deformation and UV/IR mixing,} Nucl.\ Phys.\ B {\bf 748} (2006) 473,  doi:10.1016/j.nuclphysb.2006.05.004 [hep-th/0507030].

\bibitem{Meljanac:2011cs}  S.~Meljanac, A.~Samsarov, J.~Trampetic and M.~Wohlgenannt,
  {\it Scalar field propagation in the $\phi^4$ kappa-Minkowski model,}  JHEP {\bf 12} (2011) 010,  arXiv:1111.5553 [hep-th].

\bibitem{Meljanac:2017grw}
  S.~Meljanac, S.~Mignemi, J.~Trampetic and J.~You,
 {\it Nonassociative Snyder $\phi^4$ Quantum Field Theory,}
  Phys.\ Rev.\ D {\bf 96} (2017) no.4,  045021,  doi:10.1103/PhysRevD.96.045021, [arXiv:1703.10851 [hep-th]].

\bibitem{Meljanac:2017jyk}
  S.~Meljanac, S.~Mignemi, J.~Trampetic and J.~You,
{\it UV-IR mixing in nonassociative Snyder $\phi^4$ theory,}
  Phys.\ Rev.\ D {\bf 97} (2018) no.5,  055041,  doi:10.1103/PhysRevD.97.055041, [arXiv:1711.09639 [hep-th]].

\bibitem{Cohen:1998zx}
 A.~G.~Cohen, D.~B.~Kaplan, and A.~E.~Nelson,
{\it Effective field theory, black holes, and the cosmological constant,}
 Phys.\ Rev.\ Lett.\  {\bf 82}, 4971 (1999),  [arXiv:hep-th/9803132].

\bibitem{Horvat:2010km}
  R.~Horvat, J.~Trampetic,
 {\it Constraining noncommutative field theories with holography,}
  JHEP {\bf 1101 } (2011)  112,  [arXiv:1009.2933 [hep-ph]].

\bibitem{Li:2006jja}
  M.~Li, W.~Song, Y.~Song and T.~Wang,
 {\it A Weak gravity conjecture for scalar field },
  JHEP {\bf 0705} (2007) 026, doi:10.1088/1126-6708/2007/05/026 [hep-th/0606011].

\bibitem{Huang:2006tz}
  Q.~G.~Huang and J.~H.~She,
 {\it Weak Gravity Conjecture for Noncommutative Field Theory},
  JHEP {\bf 0612} (2006) 014,  doi:10.1088/1126-6708/2006/12/014  [hep-th/0611211].

\bibitem{Palti:2017elp}
  E.~Palti,
{\it The Weak Gravity Conjecture and Scalar Fields,}
 JHEP {\bf 1708} (2017) 034,  doi:10.1007/JHEP08(2017)034  [arXiv:1705.04328 [hep-th]].

\bibitem{Lust:2017wrl}
  D.~Lust and E.~Palti,
 {\it Scalar Fields, Hierarchical UV/IR Mixing and The Weak Gravity Conjecture,}
  JHEP {\bf 1802} (2018) 040, doi:10.1007/JHEP02(2018)040  [arXiv:1709.01790 [hep-th]].

\bibitem{Craig:2019zbn}
N.~Craig and S.~Koren,
{\it IR Dynamics from UV Divergences: UV/IR Mixing, NCFT, and the Hierarchy Problem,}
JHEP \textbf{03} (2020), 037
doi:10.1007/JHEP03(2020)037
[arXiv:1909.01365 [hep-ph]].

\bibitem{Koren:2020biu}
S.~Koren,
{\it The Hierarchy Problem: From the Fundamentals to the Frontiers,}
[arXiv:2009.11870 [hep-ph]].

\bibitem{Calmet:2001na}
  X.~Calmet, B.~Jurco, P.~Schupp, J.~Wess and M.~Wohlgenannt,
  {\it The Standard model on noncommutative space-time,}
  Eur.\ Phys.\ J.\ C {\bf 23} (2002) 363
  doi:10.1007/s100520100873
  [hep-ph/0111115].

\bibitem{Aschieri:2002mc}
  P.~Aschieri, B.~Jurco, P.~Schupp and J.~Wess,
  {\it Noncommutative GUTs, standard model and C,P,T,}
  Nucl.\ Phys.\ B {\bf 651} (2003) 45
  doi:10.1016/S0550-3213(02)00937-9
  [hep-th/0205214].

\bibitem{Martin:2016zon}
  C.~P.~Martin, J.~Trampetic and J.~You,
  {\it Super Yang-Mills and $\theta$-exact Seiberg-Witten map: absence of quadratic noncommutative IR divergences,} JHEP {\bf 1605} (2016) 169,  doi:10.1007/JHEP05(2016)169, [arXiv:1602.01333 [hep-th]].

\bibitem{Martin:2016hji}
  C.~P.~Martin, J.~Trampetic and J.~You,
 {\it Equivalence of quantum field theories related by the $\theta$-exact Seiberg-Witten map,} Phys.\ Rev.\ D {\bf 94} (2016) no.4,  041703,  doi:10.1103/PhysRevD.94.041703, [arXiv:1606.03312 [hep-th]].

\bibitem{Martin:2016saw}
  C.~P.~Martin, J.~Trampetic and J.~You,
  {\it Quantum duality under the $\theta$-exact Seiberg-Witten map,}
  JHEP {\bf 1609} (2016) 052,  doi:10.1007/JHEP09(2016)052
  [arXiv:1607.01541 [hep-th]].


\bibitem{Martin:2017nhg}
  C.~P.~Martin, J.~Trampetic and J.~You,
 {\it Quantum noncommutative ABJM theory: first steps,}
  JHEP {\bf 1804} (2018) 070,  doi:10.1007/JHEP04(2018)070
  [arXiv:1711.09664 [hep-th]].

  \bibitem{DeWitt1}
  B.S. DeWitt, {\it Quantum Theory of Gravity 2. The Manifestly Covariant Theory}, Phys. Rev. 162 (1967) 1195 [INSPIRE],

\bibitem{Kallosh:1974yh}
  R.E. Kallosh, {\it The Renormalization in Nonabelian Gauge Theories}, Nucl. Phys. B 78 (1974), [INSPIRE], 
  doi:10.1016/0550-3213(74)90284-3.

\bibitem{DeWitt2}
 B.S. DeWitt, {\it A gauge invariant effective action}, in 
Oxford Conference on Quantum Gravity, 1980, NSF-ITP-80-31.

\bibitem{DeWitt3}
 B.S. DeWitt,  {\it Dynamical Theory Of Groups And Fields, 
Modern Kaluza-Klein Theories,} edited by T. Appelquist et al. 
(Addison-Wesley, Reading, MA, 1987), p.114; Relativity,
groups and topology, edited by C. DeWitt (Gordon and Breach, New York, 1965), p.725.
 

\bibitem{Grozin:2000cm}
A.~G.~Grozin,
{\it Lectures on perturbative HQET.1}, [arXiv:hep-ph/0008300 [hep-ph]].

\bibitem{Martin:2013lba}
  C.~P.~Martin,
  {\it SO(10) GUTs with large tensor representations on Noncommutative Space-time,}
  Phys.\ Rev.\ D {\bf 89} (2014) no.6,  065018
  doi:10.1103/PhysRevD.89.065018
  [arXiv:1311.2826 [hep-th]].

\bibitem{Raju:2009yx}
  S.~Raju, {\it The Noncommutative S-Matrix,}
  JHEP {\bf 0906} (2009) 005,  doi:10.1088/1126-6708/2009/06/005
  [arXiv:0903.0380 [hep-th]].

\bibitem{Huang:2010fc}
  J.~H.~Huang, R.~Huang and Y.~Jia,
  {\it Tree amplitudes of noncommutative U(N) Yang-Mills Theory,}
  J.\ Phys.\ A {\bf 44} (2011) 425401,  doi:10.1088/1751-8113/44/42/425401
  [arXiv:1009.5073 [hep-th]].

\bibitem{Arkani-Hamed:2017mur}
  N.~Arkani-Hamed, Y.~Bai, S.~He and G.~Yan,
 {\it Scattering Forms and the Positive Geometry of Kinematics, Color and the Worldsheet,} JHEP {\bf 1805} (2018) 096, doi:10.1007/JHEP05(2018)096,  [arXiv:1711.09102 [hep-th]].

\bibitem{Mizera:2019blq}
  S.~Mizera, {\it Kinematic Jacobi Identity is a Residue Theorem: Geometry of Color-Kinematics Duality for Gauge and Gravity Amplitudes,}  Phys.\ Rev.\ Lett.\  {\bf 124} (2020) no.14,  141601, doi:10.1103/PhysRevLett.124.141601,  [arXiv:1912.03397 [hep-th]].

\bibitem{Latas:2020nji}
D.~Latas, J.~Trampeti\'c and J.~You,
{\it Seiberg-Witten map Invariant Scatterings,}
Phys. Rev. D \textbf{104} (2021) no.1, 015021, doi:10.1103/PhysRevD.104.015021,
[arXiv:2012.07891 [hep-ph]].

\bibitem{Horvat:2020ycy}
  R.~Horvat, D.~Latas, J.~Trampetic and J.~You,
 {\it Light-by-Light Scattering and Spacetime Noncommutativity,}
   Phys.\ Rev.\ D {\bf 101} (2020)  095035,  doi: 10.1103/PhysRevD.101.095035,
  arXiv:2002.01829 [hep-ph].

\bibitem{Aaboud:2017bwk}
  M.~Aaboud {\it et al.} [ATLAS Collaboration],
 {\it Evidence for light-by-light scattering in heavy-ion collisions with the ATLAS detector at the LHC,}
  Nature Phys.\  {\bf 13} (2017) no.9,  852,  doi:10.1038/nphys4208
  [arXiv:1702.01625 [hep-ex]].

\bibitem{Aad:2019ock}
  G.~Aad {\it et al.} [ATLAS Collaboration],
 {\it Observation of light-by-light scattering in ultraperipheral Pb+Pb collisions with the ATLAS detector,}  Phys.\ Rev.\ Lett.\  {\bf 123} (2019) no.5,  052001, doi:10.1103/PhysRevLett.123.052001, [arXiv:1904.03536 [hep-ex]].

\end{thebibliography}
\end{document}